\documentclass[fleqn,10pt]{wlscirep}
\usepackage[utf8]{inputenc}
\usepackage[T1]{fontenc}
\usepackage{amsmath}
\usepackage{latexsym}
\usepackage{amsmath, amssymb, amscd, amsthm, amsfonts}
\usepackage{graphicx}
\usepackage{lipsum}
 \usepackage{booktabs}
 \usepackage{tablefootnote}
\usepackage{verbatim}
\usepackage{xcolor}
\usepackage{booktabs}
\usepackage{multirow}
\usepackage[export]{adjustbox}
\usepackage{booktabs}
\usepackage{threeparttable, tablefootnote}
\usepackage{xr}
\usepackage{url}

\makeatletter
\newcommand*{\addFileDependency}[1]{
  \typeout{(#1)}
  \@addtofilelist{#1}
  \IfFileExists{#1}{}{\typeout{No file #1.}}
}
\makeatother

\newcommand*{\myexternaldocument}[1]{
    \externaldocument{#1}
    \addFileDependency{#1.tex}
    \addFileDependency{#1.aux}
}
\myexternaldocument{Supplementary_Information}

\title{A Comparative Study of Compartmental Models for COVID-19 Transmission in Ontario, Canada}

\author[1]{Yuxuan Zhao}
\author[1,*]{Samuel W.K. Wong}

\affil[1]{University of Waterloo, Department of Statistics and Actuarial Science, Waterloo, N2L 3G1, Canada}

\affil[*]{samuel.wong@uwaterloo.ca}


\begin{abstract}
The number of confirmed COVID-19 cases reached over 1.3 million in Ontario, Canada by June 4, 2022. The continued spread of the virus underlying COVID-19 has been spurred by the emergence of variants since the initial outbreak in December, 2019. Much attention has thus been devoted to tracking and modelling the transmission of COVID-19. Compartmental models are commonly used to mimic epidemic transmission mechanisms and are easy to understand. Their performance in real-world settings, however, needs to be more thoroughly assessed. In this comparative study, we examine five compartmental models -- four existing ones and an extended model that we propose -- and analyze their ability to describe COVID-19 transmission in Ontario from January 2022 to June 2022. 
\end{abstract}
\begin{document}

\flushbottom
\maketitle

\thispagestyle{empty}

\section*{Introduction}

Humans have faced severe infectious diseases throughout history, some of which have been classified as worldwide pandemics \cite{novel2009emergence}, including the Spanish flu in 1917 and Hong Kong flu (H3N2) in 1968. The most recent example is the spread of the coronavirus disease 2019 (COVID-19). COVID-19 is the infectious disease caused by the novel coronavirus of severe acute respiratory syndrome (SARS-CoV-2), and the first case was detected in the Wholesale Seafood Market in Wuhan City, Hubei province, China, on December 3, 2019 \cite{https://doi.org/10.1002/jmv.25678}. The disease then spread all over the world, such that in February 2020 the World Health Organization (WHO) declared COVID-19 to be a worldwide pandemic. In a variety of ways, COVID-19 and its associated public health policies have had serious impacts on human physical and mental health since then, in many regions of the world. Up to June 2022, the cumulative confirmed cases of COVID-19 worldwide reached 530 million, and continues to increase rapidly due to the spread of variants.

Mutations in the virus underlying COVID-19 have led to a number of variants of concern, including Alpha (B.1.1.7), Beta (B.1.351), Gamma (P.1), Delta (B.1.617.2), and Omicron (B.1.1.529) \cite{who}. Omicron was the most recently detected variant in Ontario, Canada, first identified in a traveller by the Public Health Ontario laboratory on November 22, 2021 \cite{pho}. As of January 20, 2022, Omicron (and its subvariants) has become dominant and represents the majority of infections in Ontario.

Mathematical models are widely used to describe the evolution of epidemics. In particular, compartmental models are one of the most popular classes of models in epidemiology to mimic the transmission dynamics. They have played an instrumental role in tracking epidemiological trends, generating predictions, and informing decisions of policy-makers. For example, in May 2020 the British Columbia government released a management strategy for COVID-19 that heavily relied on the results of a fitted dynamic compartmental model \cite{bcgovernmentcompartmental}. Their model predicted the number of people who would require critical care under different levels of social contacts, which in turn informed the level of lockdown restrictions implemented by the government to protect the health system from being overwhelmed. 

Compartmental models divide the total population into a number of different compartments; then the flow of the population through these compartments is usually modelled via a system of differential equations. Names of compartment models are usually given in acronym form, by abbreviating the first letter of each compartment and arranging the letters according to how the population tends to flow through the different compartments \cite{garg2021mathematical}.
The earliest and simplest compartmental model was the SIR model \cite{william}. The SIR model consists of three compartments that divide the total population into susceptible (S), infected (I) and recovered (R) individuals, along with a system of three differential equations which describes the flow rate in and out of each compartment. Since then, researchers have devoted much attention towards developing extensions to the basic SIR model.

The choice of compartments, flow directions, and parameters included in these models depend mainly on the characteristics of the disease. 
We briefly overview the variety of compartmental models that have been used to describe the transmission of COVID-19. Starting from the SIR model \cite{jayatilaka2022mathematical}, the SEIR \cite{he2020seir}, SEIRD \cite{carcione2020simulation}, and SMEIHRDV \cite{compartmental} are examples of models that divide the overall population into finer compartments. Other compartmental models stratify the 
population into different groups to describe the transmission dynamics of COVID-19 in a more targeted way: e.g., 
stratification by age groups, such as the young, adults and seniors \cite{FIELDS2021e07905}; profession stratification, such as healthcare workers and others \cite{masandawa2021mathematical}; gender stratification \cite{cartocci2021compartment}. 
Some researchers combine these two approaches by setting up compartments within strata. A SEAPIR model \cite{day2020evolutionary}, which was used to model Omicron cases in British Columbia \cite{covarrnet}, stratified the susceptible population by vaccination status, i.e., vaccinated and unvaccinated. Moreover, it added an asymptomatic infection compartment to account for the high asymptomatic carriage rate of the Omicron variant \cite{garrett2022high}. Finally, a $\mathrm{SV^2(AIR)^3}$ model \cite{layton2022understanding} not only considered asymptomatic infections and vaccination status, but also added the impact of policy measures and competition between different variants.

For illustration, this paper also develops our proposed extension to the SEAPIR model. As in the SEAPIR model, we stratify the population by vaccination status and include compartments for asymptomatic (A) and pre-symptomatic (P) infections. We incorporate a time-dependent function that aggregates the policy measures that the government has imposed to reduce COVID-19 transmission. In addition, we introduce a new compartment (Q) for individuals in self-isolation, and consider the interaction between groups with different vaccination statuses in the disease transmission stage.

With the plethora of models available, the choice of a suitable one to describe epidemic transmission is therefore an important consideration. Simple models (e.g., SIR) rely on few parameters and assumptions, and thus tend to provide an oversimplified representation of reality. In contrast, a complex model will often have been designed to provide a more comprehensive description of transmission dynamics and population behaviour. However, a complex model requires a larger number of unknown parameters, which can significantly impact its performance. Unknown parameters either need to be calibrated or have their values assumed — the former can increase the variance and uncertainty of the model predictions, while the latter can introduce significant bias.

Conventional approaches for parameter calibration, such as non-linear least squares and maximum likelihood, may fail to adequately capture the uncertainties of the calibrated parameters. Their calibration results are largely dependent on the stability of the known parameters, such as recovery rate and disease incubation rate, which are often borrowed from the existing literature. Moreover, for the least squares method, the global optimum can be difficult to find when the parameter space is large, which may result in misleading inferences. As an alternative, we adopt a Bayesian approach to inference and apply Markov Chain Monte Carlo (MCMC) methods for parameter calibration in our proposed model. A Bayesian framework allows us to incorporate prior information and coherently accounts for the uncertainty of the parameters via their posterior distributions; e.g., parameters that are not well-calibrated from data will tend to have wide credible intervals.

As a specific case study, this paper focuses on modeling confirmed COVID-19 infections in Ontario from January 2022 to June 2022. This task is potentially more challenging with Omicron's prevalence, compared to the original wild-type strain. First, there are the effects of vaccination. As COVID-19 vaccines have become widely available in Ontario, most of the population of Ontario has taken a complete dose of vaccination (i.e., fully vaccinated with one or two doses of a Health Canada authorized COVID-19 vaccine), and furthermore, some have also taken a booster dose (i.e., fully vaccinated plus one additional booster dose). However, individuals who are vaccinated or have recovered from COVID-19 in the past are still likely to be infected: vaccine effectiveness against the Omicron variant exhibits a continuous and consistent decrease after injection, and vaccination provides more limited protection against symptomatic disease caused by the Omicron variant \cite{andrews2022covid}. As a result, national re-infection associated with Omicron emergence was observed in South Africa \cite{pulliam2022increased}, the United States \cite{vandegriftsars}, and Canada \cite{ribeiro2022characterisation}. Second, the limited availability of COVID-19 testing in Ontario, especially as case loads increased due to Omicron's highly transmissible nature, hinders estimation of true infection and re-infection rates. Third, Omicron is thought to have higher rates of asymptomatic infection \cite{garrett2022high}, and thus detection is more elusive. Fourth, Ontario moved through a series of reopening phases during this period \cite{ontarionewsreopening}, which has impacts on the social behavior of the population.

To the best of our knowledge, few studies have investigated whether simple models perform worse or better than complex models for describing the recent transmission dynamics of COVID-19 in Ontario. Therefore, a comparative study between models can help address this research question. This paper considers five different models: SIR \cite{jayatilaka2022mathematical}, vaccination-stratified SIR \cite{fisman2022impact}, SEIRD \cite{melo2022application}, $\mathrm{SV^2(AIR)^3}$ \cite{layton2022understanding} and our SEAPIR-extended model. We calibrate their adjustable parameters and evaluate their fits to Ontario's confirmed daily case counts. By examining their performance in this real-world setting, we gain insight into the relative strengths and shortcomings of compartmental models that range from simple to complex.

\section*{Methods}

\subsection*{Data Description}
The COVID-19 data used in this paper are obtained from Public Health Ontario. We investigate the daily confirmed COVID-19 cases from January 6 to June 4, 2022, which spans five reopening phases as determined by the Ontario government \cite{ontarionewsreopening}. The first phase is from January 6 to January 30, when the province returned to a modified Step 2 of the reopening plan \cite{ctvnews} with restrictions on social activities. The second phase is from January 31 to February 16, when the Ontario government began the process of gradually easing restrictions while maintaining protective measures \cite{ontarioreop}. The third phase is from February 17 to February 28, when the Ontario government further eased public health measures \cite{ontarionewsreopening}. The fourth phase is from March 1 to March 20, when the proof of vaccination requirement was lifted for all settings \cite{ontarioreop}. The data ends with a portion of the fifth phase from March 21 to June 4, which corresponds to the time when the Ontario government scrapped most mask mandates \cite{mask}. Daily vaccination counts were also available for this investigated time period.

As the Ontario government moved from one reopening phase to the next, the restrictions on indoor and outdoor public activities were relaxed, which led to out-of-home mobility increasing over this time period. 
These phase-to-phase changes in restrictions may be quantified via the
`Oxford Stringency Index' (denoted by $\lambda(t)$), which is an aggregate value (ranging from  0$\%$ to 100$\%$) that quantifies the ``overall impact of policy measures on workplace closures, school closures, travel bans, and vaccination requirements'' \cite{oxfordstringency}.

\begin{figure}[hbt!]
    \centering
    \includegraphics[width = 0.9\textwidth, center]{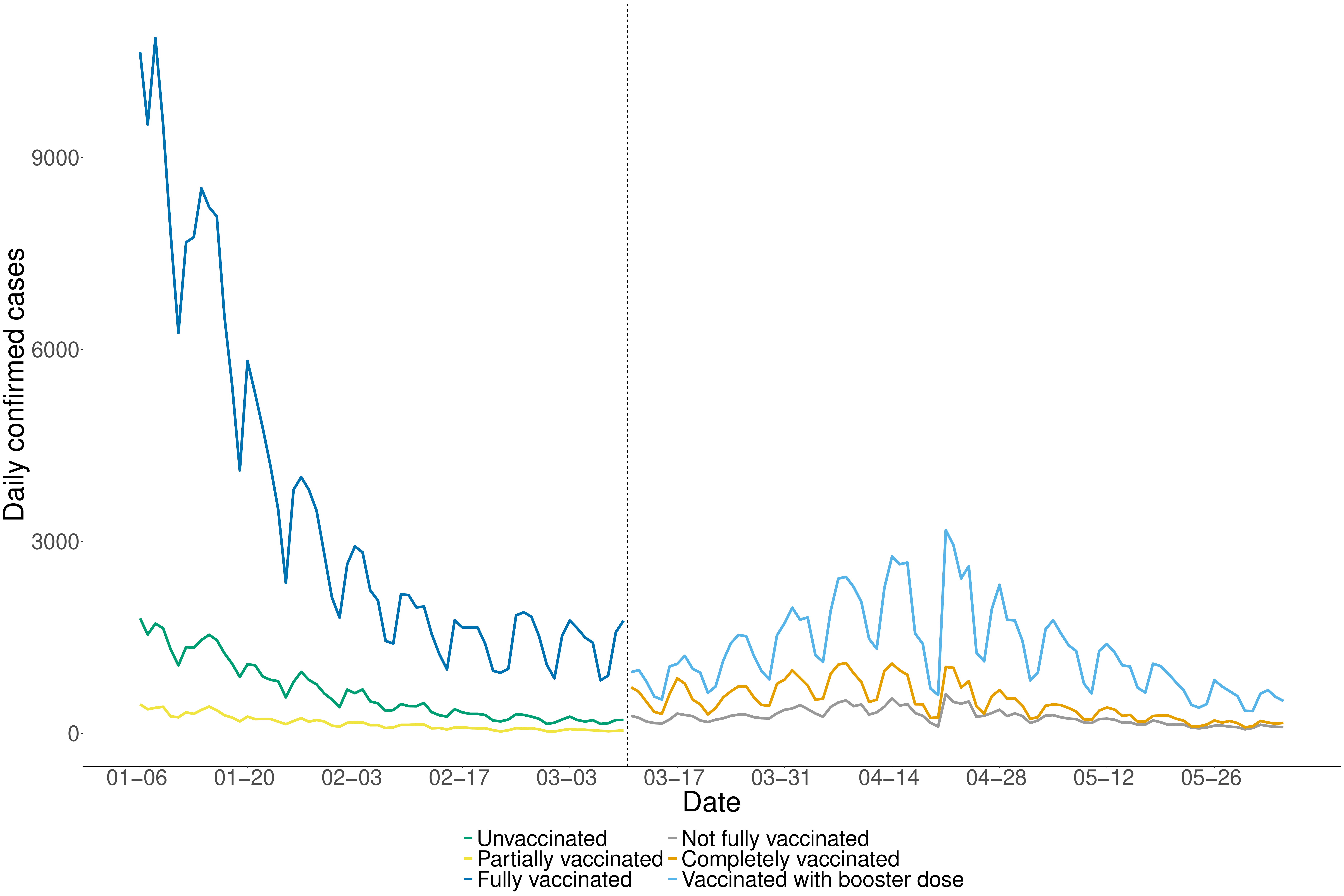}
    \caption{Daily confirmed COVID-19 cases in Ontario. The total infections are stratified by unvaccinated, partially vaccinated and fully vaccinated infection in Ontario from January 6 to March 10 (before the dashed line). From March 11 to June 4 (after the dashed line), the total infections are stratified by not fully vaccinated, completely vaccinated, and vaccinated with booster dose.}
    \label{fig:vac_status_3}
\end{figure}

The daily COVID-19 cases stratified by vaccination status shown in Figure  \ref{fig:vac_status_3} provide a more detailed look at the data \cite{cases_data}. We note that the Ontario government changed its stratification rules for reporting cases during the investigated time period. From January 6 to March 10 (before the dashed line in Figure \ref{fig:vac_status_3}), the Ontario government used three strata for reporting infections: unvaccinated, partially vaccinated, and fully vaccinated (which includes infections among both those with a completed primary series and those with an additional booster dose). From March 11 to June 4 (after the dashed line in Figure \ref{fig:vac_status_3}), the Ontario government changed the compositions of the three strata it used for reporting infections: not fully vaccinated (which included partially vaccinated and unvaccinated), completely vaccinated, and vaccinated with booster dose. Table \ref{tab:stratification_rule} shows these two different stratification rules in detail.

\begin{table}[hbt!]
\caption{Strata definitions used by Ontario government for reporting daily confirmed COVID-19 case counts, from January 2022 to June 2022.}
\resizebox{\textwidth}{!}{

\begin{tabular}{clcccc}
\hline
Date range      & Stratum name         & \multicolumn{1}{l}{Not vaccinated} & \multicolumn{1}{l}{One dose of two-dose primary series} & \multicolumn{1}{l}{Completed primary series} & \multicolumn{1}{l}{Primary series with additional booster dose} \\ \hline
\multirow{3}{*}{January 6 to March 10} & Unvaccinated                 & $\checkmark$                            &                                              &                                         &                                                    \\
                                                                                                                          & Partially vaccinated         &                                         & $\checkmark$                                 &                                         &                                                    \\
                                                                                                                          & Fully vaccinated             &                                         &                                              & $\checkmark$                            & $\checkmark$                                       \\ \hline
\multirow{3}{*}{March 11 to June 4}    & Not fully vaccinated         & $\checkmark$                            & $\checkmark$                                 &                                         &                                                    \\
                                                                                                                          & Completely vaccinated        &                                         &                                              & $\checkmark$                            &                                                    \\
                                                                                                                          & Vaccinated with booster dose &                                         &                                              &                                         & $\checkmark$                                       \\ \hline
\end{tabular}}
\label{tab:stratification_rule}
\end{table}

\subsection*{Model Descriptions}
In this section, the SIR \cite{jayatilaka2022mathematical}, vaccination-stratified SIR \cite{fisman2022impact}, SEIRD \cite{melo2022application}, $\mathrm{SV^2(AIR)^3}$ \cite{layton2022understanding}, and a new model that we call vaccination-stratified SEPAIQRD, are introduced.
\subsubsection*{SIR model}
Among others, Jayatilaka et al. \cite{jayatilaka2022mathematical} used the basic SIR model to track the spread of COVID-19. The total population size, $N$, is represented as $N = S(t) + I(t) + R(t)$, where $S(t)$, $I(t)$, and $R(t)$ respectively denote the number of susceptible, infected, and recovered individuals at time $t$.
The SIR model is governed by the following system of differential equations:
$$
\begin{cases}
\frac{\mathrm{d} S}{\mathrm{d}  t}=-\frac{\beta S I}{N}\\
\frac{\mathrm{d}  I}{\mathrm{d}  t}=\frac{\beta S I}{N}-\gamma I \\
\frac{\mathrm{d}  R}{\mathrm{d}  t}=\gamma I,
\end{cases}
$$
where $\gamma$ and $\beta$ are the two parameters to be calibrated. The parameter $\beta$, as seen in the first two equations, governs the rate at which individuals move from the $S$ to the $I$ compartment; it is called the disease transmission rate, representing the average number of susceptible individuals in the $S$ compartment that an contagious individual in the $I$ compartment infects in a day. The parameter $\gamma$, as seen in the last two equations, governs the rate at which individuals move from the $I$ to the $R$ compartment; it is called the removal rate, representing the probability per day that an $I$-individual transits to $R$ (which can encompass both recovered and deceased individuals). Thus, the average duration of infection under this model is $1/\gamma$. Note that the last equation implies that $R$ is an absorbing state, since individuals can no longer leave once they enter this compartment (e.g., reinfections are not possible in this model). The unknown parameters $\beta$ and $\gamma$ can be calibrated by minimizing the sum of squared errors (SSE) between the model-fitted daily case counts and actual case counts.

\subsubsection*{Vaccination-stratified SIR model}
Fisman et al. \cite{fisman2022impact} proposed a modified SIR model by stratifying the population into vaccinated and unvaccinated groups. They also considered the impact of interaction (or mixing) between the vaccinated and unvaccinated sub-populations on COVID-19 transmission, by introducing the parameters $f_{ij}$ for the fraction of contacts among individuals in the i$^\mathrm{th}$ group (i.e., vaccinated or unvaccinated) with those in the j$^\mathrm{th}$ group. Immunity from vaccination (when effective) is assumed to be permanent. The parameters $\gamma$ and $\beta$ have the same interpretation as in the basic SIR model.

The following differential equations govern this model:
$$
\begin{cases}
\frac{\mathrm{d} S_i}{\mathrm{d}  t}=-\beta \sum f_{ij} \frac{I_j}{N_j}\\ 
\frac{\mathrm{d}  I_i}{\mathrm{d}  t}=\beta \sum f_{ij} \frac{I_j}{N_j}-\gamma I_i\\
\frac{\mathrm{d}  R_i}{\mathrm{d}  t}=\gamma I_i,
\end{cases}
$$
where $N_i= S_i(t) + I_i(t) + R_i(t)$ is the subpopulation size of the i$^\mathrm{th}$ group.

The authors mainly obtained their parameters from other literature and numerically solved the differential equations with predetermined initial conditions. Therefore, their parameters are not designed for calibration from data, which may adversely impact model performance when applied to situations where their assumptions do not hold.

\subsubsection*{SEIRD model}

Melo \cite{melo2022application} proposed an SEIRD model to provide a fuller description of COVID-19 progression, which divides the population into finer compartments. Susceptible individuals $S$ first move to the exposed compartment $E$ with disease transmission rate $\beta$, rather than directly moving to $I$. After the disease incubation period (an average of $1/\gamma$ days), exposed individuals will transit into the $I$ compartment. 
Infected individuals will then either move to the recovered $R$ compartment (with a rate of $\mu$) or the dead $D$ compartment (with a rate of $\rho$).
Their governing system of differential equations is shown as follows:
$$
\begin{cases}
\frac{d S}{d t}  =-\frac{\beta S I}{N}\\
\frac{d E}{d t} =\frac{\beta S I}{N}-\gamma E\\
\frac{d I}{d t}  =\gamma E-\rho I-\mu I\\
\frac{d R}{d t} =\mu I\\
\frac{d D}{d t}  =\rho I. 
\end{cases}
$$
As in the SIR model, the unknown parameters $\beta, \gamma, \rho,\mu$ can be calibrated by minimizing the SSE between the model-fitted daily confirmed case counts and the actual ones.

\subsubsection*{$\mathrm{SV^2(AIR)^3}$ model}
Layton and Sadria \cite{layton2022understanding} introduced the $\mathrm{SV^2(AIR)^3}$ model to provide a more comprehensive description of COVID-19 epidemic progression in Ontario. The authors included additional parameters that measured the impact of waning immunity, vaccine effectiveness, and policy measures that restrict public activities. The quantitative values of measuring the policy strictness are equivalent to the previously introduced Oxford Stringency Index. The compartmental setup considered two vaccine types (hence $V^2$), asymptomatic infections (A), and competition among the three main variant types as of Fall 2021, i.e., wild, Alpha, and Delta. The authors also modeled the potential spread of a hypothetical new-emerging variant.

The model parameters that describe the clinical characteristics of the COVID-19 variants are obtained from published studies. Other parameters related to the demographics and social behaviours of the Ontario population are obtained from published provincial statistics. In total, the model has 69 parameters. To calibrate the model, we update nine parameters pertaining to their new-emerging variant to mimic the characteristics of the actual Omicron variant, including higher values for Omicron's transmission rate and fraction of asymptomatic infection. We also use the actual values of the Oxford Stringency Index during our investigated period, which serve as scaling factors in the model. Table \ref{tab:para_sav_alpha} and \ref{tab: para_sav_delta} in the Supplementary Information respectively show the model parameters with respect to wild-type, Alpha-type, Delta-type, and our updated Omicron-type variants.

\subsubsection*{Vaccination-stratified SEPAIQRD model}
We introduce an extension of the SEAPIR model to describe the dynamic mechanisms of COVID-19 transmission in Ontario over the investigated period, which we call the vaccination-stratified SEPAIQRD model. A summary of its key features is as follows:  more compartments are added to reflect the situation in Ontario; the population is stratified by the four vaccination statuses as defined by Ontario; migration between susceptible compartments of the different vaccination statuses occurs, according to the daily reported vaccination counts.

The COVID-19 data released by Public Health Ontario of confirmed COVID-19 cases are split into three strata up to and including March 10, as presented in the Data Description section. During this period, the `fully vaccinated' infections counted both completely vaccinated and vaccinated with booster dose infections.  For simplicity, we further split these `fully vaccinated' infections according to the daily-updated proportion of completely vaccinated and vaccinated with booster dose populations in Ontario. After March 10, we further split the `not fully vaccinated' infections according to the daily-updated proportion of unvaccinated population and partially vaccinated population in Ontario. Figure \ref{fig:str} in the Supplementary Information plots the case counts stratified by the four vaccination statuses after this processing step. In the following description, the superscript index $i$ for $i=1,2,3,4$ will respectively denote the unvaccinated, partially vaccinated, completely vaccinated, and vaccinated with booster dose populations. We let $N^i$ denote the size for each of these populations. The flow of susceptible individuals with these four vaccination statuses will be tracked in the model using parallel compartments.

Susceptible individuals (in the `$S^i$' compartment) can move to the exposed compartment (denoted as `$E^i$') when in contact with contagious individuals. We let $\beta_k^{ij}$ denote the transmission rate from the contagious compartment $k$ in the $i^{\mathrm{th}}$ group to the susceptible individuals in the $j^{\mathrm{th}}$ group. The construction of the disease transmission matrix that governs such interactions follows a previous approach \cite{FIELDS2021e07905} and is described in Section B.2 of the Supplementary Information, with Table \ref{tab:contact} showing the contact matrix given different vaccination statuses. Tables \ref{tab:contact_IP} and \ref{tab:contact_IA} in the Supplementary Information show the transmission matrix of different contagious compartments.  These disease transmission rates will be scaled multiplicatively by $1-\lambda(t)$, which quantifies the impact of policy measures via the Oxford Stringency Index.

\begin{figure}[hbt!]
     \centering
     \includegraphics[width = 1\textwidth]{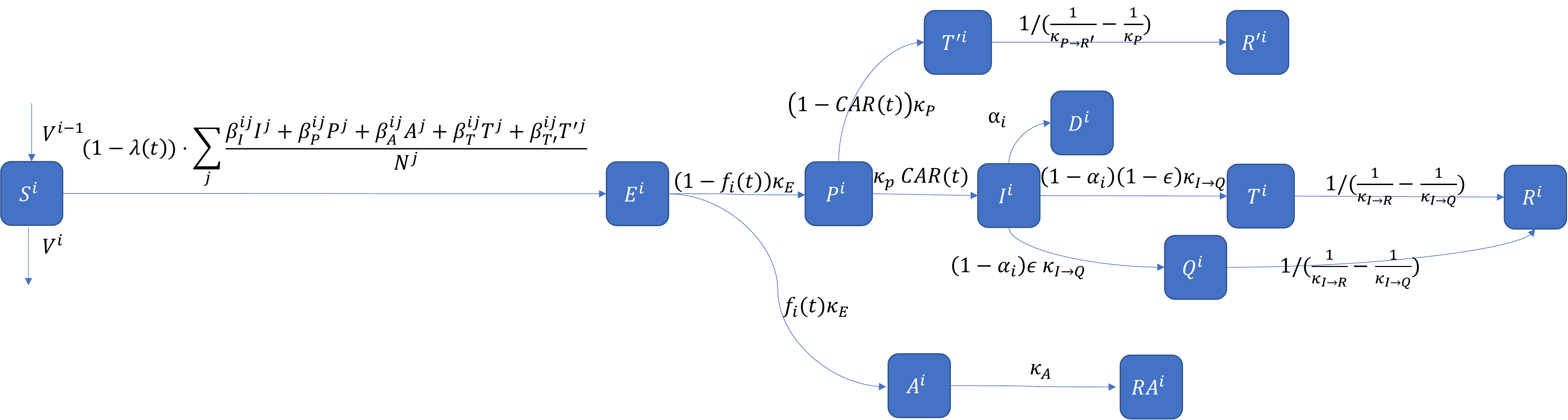}
     \caption{Schematic diagram of vaccination-stratified SEPAIQRD model in the i$^\mathrm{th}$ vaccination group. The arrows indicating flux in and flux out of the $S^i$ compartment will depend on vaccination status: there is no $V^{i-1}$ for the unvaccinated group and no $V^i$ for the vaccinated with booster group.}
     \label{fig:proposed_model_diagram}
 \end{figure}

After exposure, asymptomatic individuals are assumed to follow the flow $E^i \to A^i$ (asymptomatic) $\to RA^i$ (recovered asymptomatic). Those with mild to severe symptoms follow the flow $E^i \to P^i$ (pre-symptomatic); then after the disease incubation period, they either recover without testing ($P^i \to R'^{i}$, e.g., mild symptoms) or are documented by the Ontario government as confirmed cases ($P^i \to I^i$, e.g., more serious symptoms). Finally, individuals with confirmed cases follow one of three flows: $I^i \to D^i$ (death); $I^i \to Q^i$ (quarantined) $\to R^i$ for those who self-isolate and then recover; $I^i \to R^i$ for those who recover without self-isolation. These compartments and flows are all illustrated in the overall schematic of the model in Figure \ref{fig:proposed_model_diagram}.

The flows in Figure \ref{fig:proposed_model_diagram} are governed by a number of fixed and time-varying parameters. The fixed $\kappa$ parameters \cite{covarrnet,FIELDS2021e07905} are various transition rates; e.g., $\kappa_E$ governs the $E^i \to P^i$ transition rate, with the interpretation that an individual spends an average of $1/\kappa_E$ days in the $E^i$ compartment. The death rate is $\alpha_i$, whose value depends on the vaccination status \cite{mortality}.  The fixed $\epsilon$ parameter is the proportion of infected individuals who comply with self-isolation after testing positive. Table \ref{tab:parameters_proposed} in the Supplementary Information lists the values of these fixed parameters.
Next, the time-varying parameter $f_i(t)$ is interpreted as the probability of asymptomatic infection; it is treated as unknown and will be calibrated from data for each vaccination status and reopening phase. Finally, the time-varying `case ascertainment rate', denoted as $CAR(t)$, is interpreted as the proportion of symptomatic infections that are documented by the Ontario government as confirmed cases; it is also treated as unknown and will be calibrated from data for each reopening phase.  Note that $T'^i$ and $T^i$ in Figure \ref{fig:proposed_model_diagram} are intermediate compartments set up so that the parameters $CAR(t)$, $f_i(t)$, and $\epsilon$ can be interpreted as the proportion of flux-out from the preceding compartment. Table \ref{tab:para_def} summarizes all of the model parameters and their corresponding definitions. 

The final element of the model is the flow of people who migrate between vaccination statuses during the investigated period. We let $V^1$, $V^2$, and $V^3$ respectively denote the number of individuals taking first, second, and booster vaccine doses, represented as daily counts reported by the Ontario government. These daily counts of individuals who get vaccinated govern the flows $S^1 \to S^2$, $S^2 \to S^3$, and $S^3 \to S^4$, as indicated in Figure \ref{fig:proposed_model_diagram}. The corresponding population sizes $N^i$ are also updated daily based on these counts.

The overall model incorporates certain assumptions, which we now state explicitly. We assume that the cases reported by Public Health Ontario units are symptomatic cases. We expect this assumption to be reasonable, since the Ontario government decided to ``limit eligibility for publicly funded PCR tests to high-risk individuals who are symptomatic beginning from December 31 (2021)'' \cite{covidtest}. This policy was maintained throughout the investigated time period. We assume that asymptomatic and mild cases are not tested and therefore do not self-isolate. We assume that the rate at which Ontario government documents confirmed cases and asymptomatic infection rate are constant within one phase, and allowed to change between phases. Moreover, the model allows the asymptomatic infection rate to differ by vaccination status and by reopening phase, which will be calibrated from data.
 
The full system of differential equations, that corresponds to Figure \ref{fig:my_model} and incorporates the above considerations, is provided in Section B.4 of the Supplementary Information.

\begin{table}[hbt!]
\caption{Definitions of model parameters in vaccination-stratified SEPAIQRD model}
\centering
\resizebox{\textwidth}{!}{
\begin{tabular}{@{}llll@{}}
\toprule
Symbol & Definition   & Reference \\ \midrule

    $\beta^{ij}_k$   &  Transmission rate from contagious compartment $k$ in $i^{\mathrm{th}}$ group to susceptible in $j^{\mathrm{th}}$ group        &      \cite{FIELDS2021e07905}                  \\
    
    $\lambda(t)$   &    Oxford Stringency Index        &  \cite{oxfordstringency}                    \\
   $f_i(t)$   &    Phase-dependent asymptomatic infection rate at $i^{\mathrm{th}}$ vaccinated group        &  To be estimated from data            \\ 
    $\kappa_E$   &  Transition rate from being exposed to being pre-symptomatically infected          &   \cite{covarrnet}                  \\
   $\kappa_A$   &  Recovery rate from being asymptomatically infected          &           \cite{covarrnet}         \\ 
      $CAR(t)$   &  Phase-dependent case ascertainment rate          &     To be estimated from data           \\ 
         $\kappa_P$   & Transition rate from being pre-symptomatically infected to being infected with symptoms               &    \cite{covarrnet}         \\ 
           $\kappa_{P\to R'}$   &  Recovery rate directly from being pre-symptomatically infected        &     \cite{covarrnet}                   \\  
    $\alpha_i$   &   Death rate of $i^{\mathrm{th}}$ vaccinated population       &        \cite{mortality}                  \\    
    $\epsilon$   &   	Proportion of compliance with isolation        &                  Estimated       \\    
       $\kappa_{I\to R}$   &   Recovery rate from being symptomatically infected         &          \cite{covarrnet}            \\   
          $\kappa_{I\to Q}$   &  Isolation delay                         &      \cite{FIELDS2021e07905}      \\  
                                \bottomrule
\end{tabular}}
\label{tab:para_def}
\end{table}

\subsection*{Parameter calibration for vaccination-stratified SEPAIQRD model}
 
Our model has two unknown phase-dependent parameters that need to be calibrated based on data, namely $f^i(t)$ and $CAR(t)$. We take a Bayesian approach to inference and apply MCMC methods for parameter calibration. Briefly, the key idea of Bayesian inference is to incorporate prior information or beliefs concerning the unknown parameters with the likelihood of the observed data, to generate a posterior distribution for the unknown parameters. Mathematically, for parameters $\theta$ and data $y$, along with the prior distribution $\pi(\theta)$ and the likelihood function $P(y|\theta)$, by Bayes's Theorem, the posterior distribution $P(\theta|y)$ (up to a multiplicative constant) is given by
\begin{equation}
       P(\theta|y) \propto  P(y|\theta) \pi(\theta).
       \label{bayes}
\end{equation}
When closed-form analysis of the posterior distribution is not possible, MCMC methods are often used to generate samples from $P(\theta|y)$. To facilitate MCMC sampling, we convert $CAR(t)$ and $f_i(t)$ from their $[0,1]$ scale to the real numbers via a logit transformation. We let $CAR(t)$ and $f_i(t)$ on the logit scale be denoted as $\mathcal{L}(CAR(t))$ and $\mathcal{L}(f_i(t))$, where $\mathcal{L}(x) = \log(\frac{x}{1-x})$.

Given initial conditions for each compartment and a set of values for the phase-dependent $f^i(t)$ and $CAR(t)$, running the numerical ODE solver produces deterministic trajectories of each compartment.  Using the hat symbol to denote the numerical solution, the number of new daily confirmed cases in the i$^{\mathrm{th}}$ vaccination status on the t$^{\mathrm{th}}$ day can be expressed as $\Delta\widehat{I^i}(t) := \widehat{I^i}(t)+\widehat{R^i}(t)+\widehat{D^i}(t)+\widehat{Q^i}(t)-(\widehat{I^i}(t-1)+\widehat{R^i}(t-1)+\widehat{D^i}(t-1)+\widehat{Q^i}(t-1))$. 
We denote the actual daily confirmed case counts on the t$^\mathrm{th}$ day as $\Delta I^i(t)$. To create a probabilistic link between the ODE solution and the actual data, we assume that these case counts have a negative binomial likelihood:
$$
    \Delta I^i(t)|\mathcal{L}(f_i(t)), \mathcal{L}(CAR(t)),\phi^i(t) \sim \mathrm{NegBin}(\Delta \widehat{I}^i(t),\phi^i(t)),
$$
where $\phi^i(t)$ is the phase-dependent parameter that accounts for overdispersion in the i$^\mathrm{th}$ group, i.e., conditional on the ODE solution, we assume independence of the negative binomial across different days \cite{mcgregor2021assessing}.

To complete the Bayesian model, prior distributions need to be specified for the unknown phase-dependent parameters $\mathcal{L}(f_i(t))$, $\mathcal{L}(CAR(t))$, and the overdispersion parameter $\phi^i(t)$. These are chosen to be weakly informative: we use them to encode some \textit{a priori} beliefs and knowledge, while letting the data likelihood be the main contributor to the posterior. First, we believe \textit{a priori} that vaccinated individuals are more likely to be asymptomatically infected, as it is commonly accepted that mild or severe COVID-19 symptoms are reduced by vaccinations and boosters \cite{andrews2022duration}. Second, the immunity conferred by vaccination wanes over time, especially against Omicron \cite{andrews2022covid}; with most booster doses in Ontario being administered in early 2022, their overall effect is expected to wane over the investigated period. These considerations are encoded by ordering the prior means for $\mathcal{L}(f_i(t))$ according to vaccination status and phase.

Third, the daily case counts are highest in January and decrease over our investigated period, and the Ontario government may more readily document cases when daily infections are lower. Thus, the prior means for $\mathcal{L}(CAR(t))$ are set to increase with reopening phase. Due to our uncertainties about these unknown parameters, a relatively large standard deviation of the priors is chosen, so that the posterior distributions will be primarily informed by the data.
The full list of priors is shown in Table \ref{tab:prior_parameters} in the Supplementary Information for $\mathcal{L}(f_i^j)$,
$\mathcal{L}(CAR^j)$, and $(\phi_i^j)^{-1}$, for vaccination status $i \in\{ 1,2,3,4\}$ and reopening phase $j \in \{1,2,3,4,5\}$. As seen in Figure \ref{fig:my_model_para}, the prior densities (blue) chosen are relatively flat, indicating they encode some knowledge without strongly contributing to the posterior.

Multiplying the likelihood and prior (Equation \ref{bayes}) yields the posterior distribution of the unknown parameters $CAR^j,\phi_i^j,f_i^j$: 
$$
\begin{aligned}
    P(\mathcal{L}(\boldsymbol{CAR}),(\boldsymbol{\phi})^{-1},\mathcal{L}(\boldsymbol{f})|\Delta \boldsymbol{I})&\propto\prod_{t = 1}^{150} \prod_{i=1}^{4}\prod_{j=1}^{5} P(\Delta I^i(t)|\mathcal{L}(CAR^j),(\phi_i^j)^{-1},\mathcal{L}(f_i^j)) \\
    &\cdot \pi((\phi_i^j)^{-1}) \cdot \pi(\mathcal{L}(CAR^j))\cdot \pi(\mathcal{L}(f_i^j)),
    \end{aligned}
$$
where $\boldsymbol{CAR}, \boldsymbol{\phi}, \boldsymbol{f},$ and $\Delta\boldsymbol{I}$ are the concatenated vector forms of $CAR^j, \phi^j_i, f^j_i,$ and $\Delta I^i(t)$, with $i \in \left\{1,2,3,4\right\}, j \in \left\{1,2,3,4,5\right\}$.

To obtain samples from the posterior distribution of the parameters, we use Stan 2.21.5 \cite{stan} and R 4.1.1, running 2000 MCMC iterations and four chains. Let $\Delta\boldsymbol{I}$ be the concatenated vector form of the daily case counts $\Delta I^i(t)$, and $\Delta\boldsymbol{\widehat{I}}_{pred}$ the corresponding model estimates. The posterior distribution of $\Delta\boldsymbol{\widehat{I}}_{pred}$ is then approximated using the MCMC samples via
$$
    P(\Delta\boldsymbol{\widehat{I}}_{pred}|\Delta\boldsymbol{ I}) = \int P(\Delta\boldsymbol{\widehat{I}}_{pred} |\boldsymbol{\theta}) P(\boldsymbol{\theta}|\Delta \boldsymbol{I})\mathrm{d}\boldsymbol{\theta} \approx \frac{1}{K} \sum_{k=1}^K P(\Delta\boldsymbol{\widehat{I}}_{pred} |\boldsymbol{\theta^{(k)}}),
$$
where $\boldsymbol{\theta} = (\mathcal{L}(\boldsymbol{CAR}),(\boldsymbol{\phi})^{-1},\mathcal{L}(\boldsymbol{f}))$ and $\boldsymbol{\theta^{(1)}}, \ldots, \boldsymbol{\theta^{(K)}}$ are the MCMC samples of $\boldsymbol{\theta}$.  We treat the posterior mean as the model-fitted daily case counts. Further, credible intervals can be easily obtained in the Bayesian framework, e.g., we take the 0.025 and 0.975 quantiles of the posterior distribution for $\Delta\boldsymbol{\widehat{I}}_{pred}$ (based on the MCMC samples) to form 95$\%$ credible intervals.

We noted that our investigated period is divided into several reopening phases. When the Ontario government moved from one reopening phase to another, the Oxford Stringency Index will immediately reflect the change, whereas the actual social behaviour in response to the change is likely to be more gradual. Jump discontinuities from the piece-wise function $\lambda(t)$ and the phase-dependent parameters ($CAR^j$ and $f^j_i$) would lead to corresponding jumps in estimated case counts. Locally weighted linear regression (LOESS) \cite{little2013oxford} is a nonparametric technique that is useful for estimating a smoothed curve, e.g., in volatile time series data. This technique has been used to smooth SIR model predictions on the total number of deaths, in the presence of a time-varying case ascertainment rate \cite{deo2021new}. Here, we also apply LOESS with automatic bandwidth and span selection to our fitted daily case counts and credible boundaries. Another technique could be to use a smooth linear function to interpolate $\lambda(t)$ over a 1-week period after the start of a new phase \cite{mcgregor2021assessing}; however, it is less applicable here due to the presence of other phase-dependent parameters ($CAR^j,f_i^j$).
 
\subsection*{Initial Conditions for the Models} 
The total population size is a key input to each of the compartmental models described. We set $N$=14,051,980, according to the total population in Ontario as recorded in the Canadian census  \cite{canadapopulation}.
The true initial conditions for the different compartments are generally unknown, so an estimation procedure is needed. The number of active infections on a given day (e.g., corresponding to compartment $I$) is estimated by the number of hospitalized patients with COVID-19 divided by the 1.9\% hospitalization rate of COVID-19 \cite{hospitalization}. As of January 5, 2022, the number of confirmed COVID-19 cases in Ontario was approximately 840,000 \cite{cumulative_cases_jan}.
However, due to the high re-infection rate of the Omicron variant, previously infected individuals were still likely to be susceptible. Thus, for simplicity, we only used confirmed case counts in early January to set initial conditions for the remaining compartments. Specifically, with Omicron having an average recovery time of five days, we used the case count on January 1 to set the initial size of the recovered compartment. Since 840,000 represents about only 6\% of the Ontario population, this modeling choice will only have a small effect on the total number of remaining susceptible individuals, and thus compartmental model dynamics. Similar reasoning is used to obtain initial conditions for quarantined, exposed, and pre-symptomatic compartments (where applicable to the model), based on the infection counts recorded on January 5, January 11 and January 8, respectively. 

\section*{Results and Discussion}

This section presents the results of our comparative study, where the five compartmental models are fitted to the Ontario COVID-19 data over our investigated period. After calibrating the relevant parameters in each model to these data as described in the Methods section, the numerical solution of each model is computed from January 6 to June 4, 2022. These model trajectories are compared to the actual daily case counts to assess their goodness-of-fit to the data.
For models that encode fewer vaccination statuses than the stratification rules for case counts used by the Ontario government, the proportion of sub-populations with different vaccination statuses will be used to allocate the estimated case counts. This is to ensure that all of the model fits can be fairly compared to the ground truth provided by the Ontario government. 

\subsection*{Assessing the Model Fits}

\begin{figure}[hbt!]
    \centering
    \includegraphics[width = 1\textwidth]{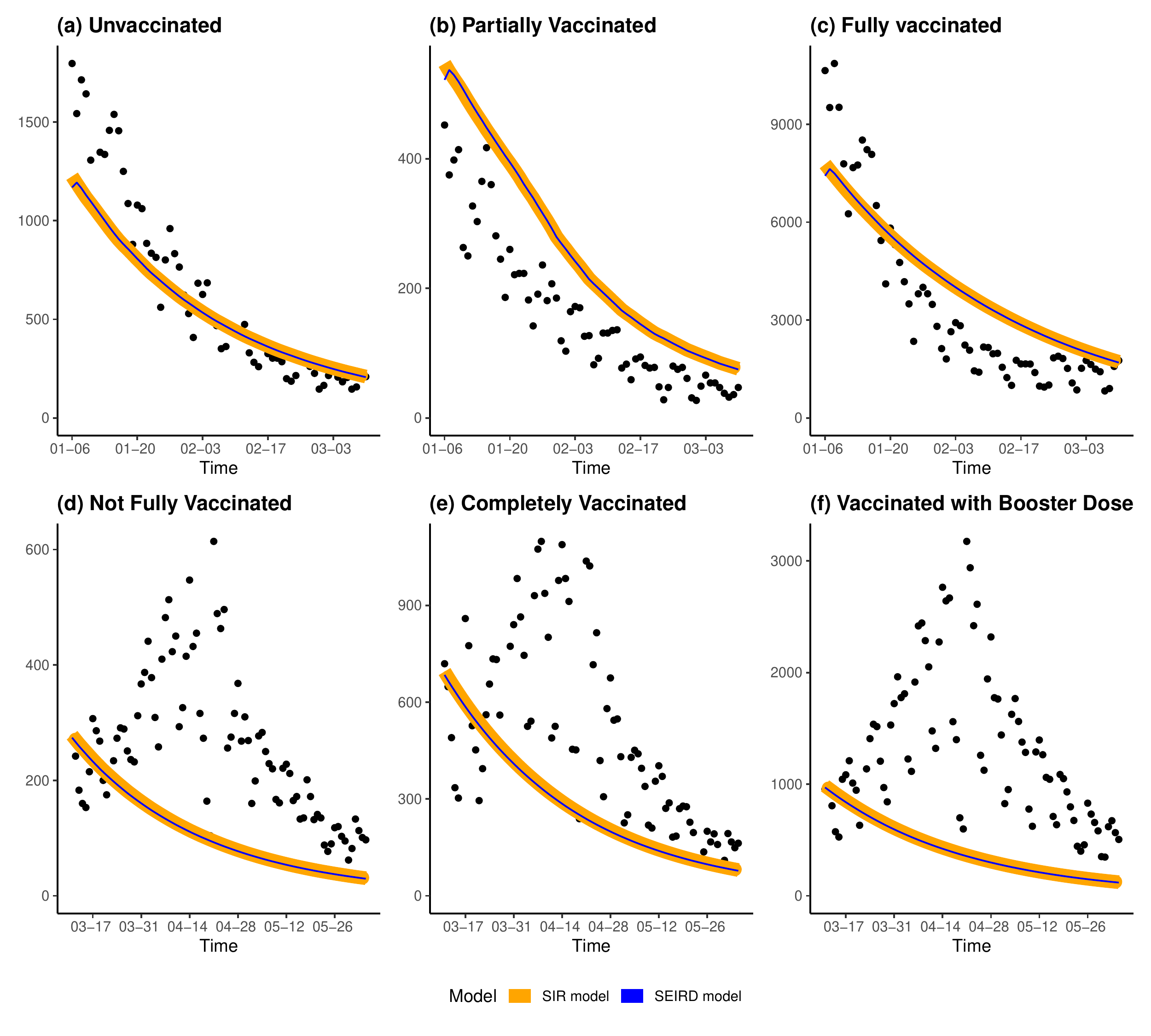}
    \caption{Trajectories of the calibrated SIR (orange lines) and SEIRD (blue lines) models over our investigated period. Black points plot the actual confirmed daily case counts. The panels correspond to the six strata definitions used by the Ontario government in Table \ref{tab:stratification_rule}.  The top panels (\textbf{a-c}) correspond to the period from January 6 to March 10, while the bottom panels (\textbf{d-f}) plot March 11 to June 4.}
    \label{fig:sir_seird}
\end{figure}

\begin{figure}[hbt!]
    \centering
    \includegraphics[width = 1\textwidth]{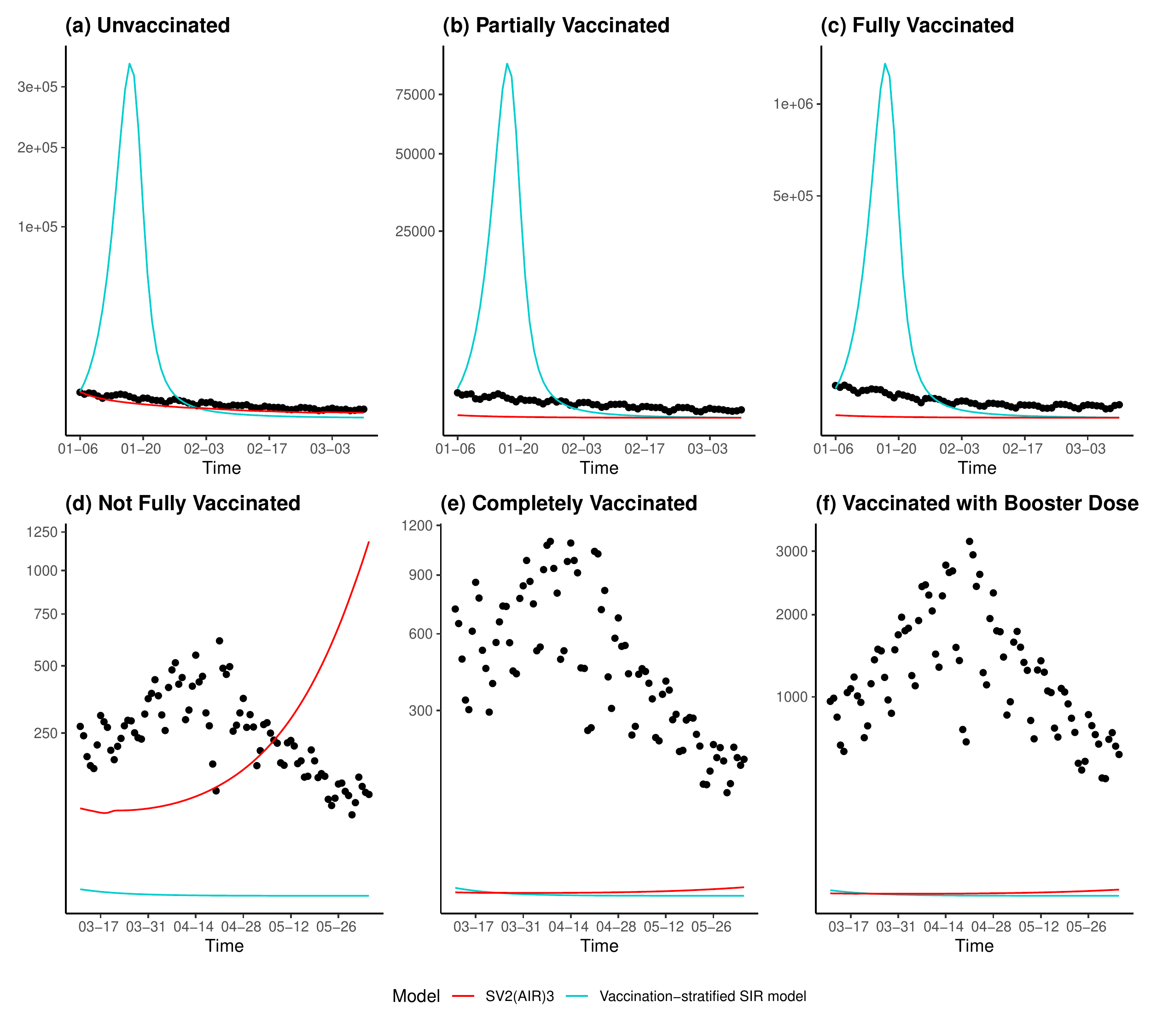}
    \caption{Trajectories of vaccination-stratified SIR model (teal lines) and Omicron-calibrated $\mathrm{SV^2(AIR)^3}$ model (red lines) over our investigated period. Black points plot the actual confirmed daily case counts. The panels correspond to the six strata definitions used by the Ontario government in Table \ref{tab:stratification_rule}.  The top panels (\textbf{a-c}) correspond to the period from January 6 to March 10, while the bottom panels (\textbf{d-f}) plot March 11 to June 4.}
    \label{fig:layton_cmaj}
\end{figure}

\begin{figure}
    \centering
    \includegraphics[width = 1\textwidth]{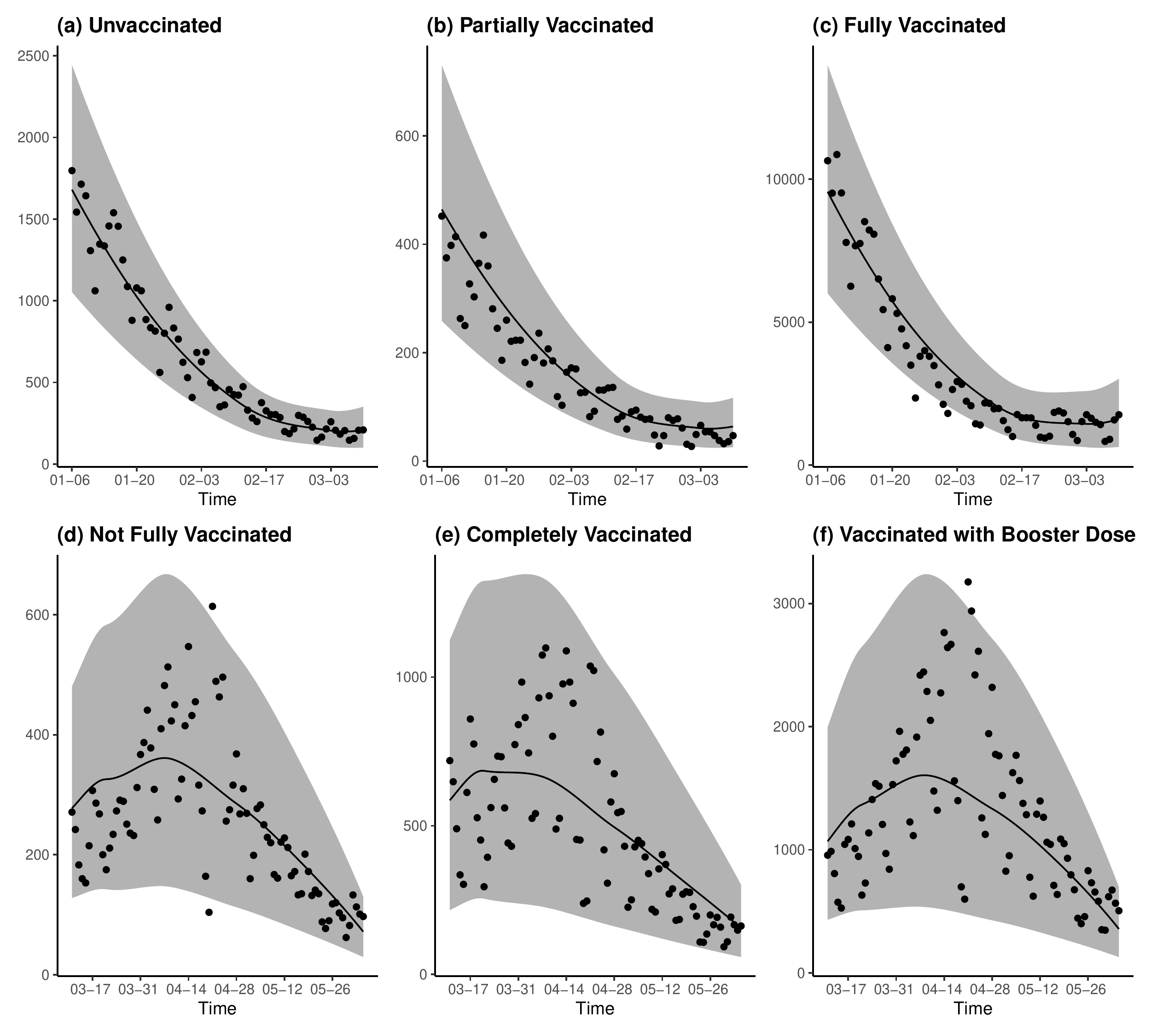}
    \caption{Trajectories of proposed vaccination-stratified SEPAIQRD model (black lines) over our investigated time period. Black dots are the actual daily confirmed new cases. The grey-shaded regions are the 95$\%$ credible bands of the trajectories. Black solid lines are the calibrated daily confirmed new cases. The panels correspond to the six strata definitions used by the Ontario government in Table \ref{tab:stratification_rule}.  The top panels (\textbf{a-c}) correspond to the period from January 6 to March 10, while the bottom panels (\textbf{d-f}) plot March 11 to June 4.}
    \label{fig:my_model}
\end{figure}

We first present graphical summaries of the model fits, by overlaying the actual confirmed daily case counts on the fitted model trajectories. The fits are shown according to the six strata definitions used by the Ontario government for reporting cases in Table \ref{tab:stratification_rule}. The trajectories of the calibrated SIR and SEIRD models are plotted in Figure \ref{fig:sir_seird}, while the trajectories of the vaccination-stratified SIR model and Omicron-calibrated $\mathrm{SV^2(AIR)^3}$ model are plotted in Figure \ref{fig:layton_cmaj}. Finally, the trajectories of the calibrated vaccination-stratified SEPAIQRD model are plotted in Figure \ref{fig:my_model}, with the grey bands representing the credible region with 95\% probability under the Bayesian posterior.

A corresponding quantitative measure of model performance can be provided by the root mean squared error (RMSE) between the model-estimated daily confirmed case counts and the actual case counts. We first compute the RMSEs for each model based on the total daily case counts (regardless of vaccination status) over the entire investigated period, as shown in Table \ref{tab:rmse_150}.  Then, we compute the RMSEs of the fitted case counts according to Ontario's stratification rules:  Table \ref{tab:rmse_64} shows the results for January 6 to March 10, and Table \ref{tab:rmse_86} shows the results for March 11 to June 4.

\begin{table}[hbt!]
\centering
\caption{RMSEs of the five models for estimating the total daily case counts in Ontario from January 6 to June 4.}

\begin{tabular}{@{}lr@{}}
\toprule
Model                                           & \multicolumn{1}{c}{RMSE} \\ \midrule
SIR model                                       & 1610.25                              \\
Vaccination-stratified SIR model                & 272840.60                          \\
SEIRD model                                     & 1613.71                             \\
$\mathrm{SV^2(AIR)^3}$ model                    & 3557.78                       \\
Vaccination-stratified SEPAIQRD model &  840.08                           \\ 
\bottomrule
\end{tabular}
\label{tab:rmse_150}
\end{table}

\begin{table}[hbt!]
\caption{RMSEs of the five models for estimating the stratified daily case counts in Ontario from January 6 to March 10.}
\centering
\resizebox{0.8\textwidth}{!}{
\begin{tabular}{@{}lrrr@{}}
\toprule
\multicolumn{1}{c}{\multirow{2}{*}{Model}}      & \multicolumn{3}{c}{RMSE in strata}                                                                               \\ \cmidrule(l){2-4} 
\multicolumn{1}{c}{}                            & \multicolumn{1}{l}{Unvaccinated} & \multicolumn{1}{l}{Partially Vaccinated} & \multicolumn{1}{l}{Fully Vaccinated} \\ \midrule
SIR model                                       &218.76                       & 108.62                                 & 1437.58                              \\
Vaccination-stratified SIR model                & 84109.78                    &22078.12                                & 311505.7                        \\
SEIRD model                                     & 218.20                       & 108.99                                 & 1430.15                              \\
$\mathrm{SV^2(AIR)^3}$ model                    & 340.76                    & 193.26                           & 4363.23                       \\

Vaccination-stratified SEPAIQRD model&      110.06                    &          43.36                          &      796.45                     \\
\bottomrule
\end{tabular}}

\label{tab:rmse_64}
\end{table}

\begin{table}[hbt!]
\caption{RMSEs of the five models for estimating the stratified daily case counts in Ontario from March 11 to June 4.}
\centering
\resizebox{\textwidth}{!}{
\begin{tabular}{@{}lrrr@{}}
\toprule
\multicolumn{1}{c}{\multirow{2}{*}{Model}}      & \multicolumn{3}{c}{RMSE in strata}                                                                               \\ \cmidrule(l){2-4} 
\multicolumn{1}{c}{}                            & \multicolumn{1}{l}{Not Fully Vaccinated} & \multicolumn{1}{l}{Completely Vaccinated} & \multicolumn{1}{l}{Fully Vaccinated with Booster Dose} \\ \midrule
SIR model                                       &189.20                         & 323.36                                  & 1116.02                       \\
Vaccination-stratified SIR model                & 283.56                    &567.14                        &1458.50                     \\
SEIRD model                                     & 190.80                      & 327.00                                  & 1122.39                              \\
$\mathrm{SV^2(AIR)^3}$ model                    & 369.98                    & 567.08                          & 1458.37                     \\

Vaccination-stratified SEPAIQRD model &     85.92                   &          197.15                        &    523.49                    \\ 
\bottomrule
\end{tabular}}
\label{tab:rmse_86}
\end{table}

The graphical and RMSE summaries indicate that none of the compartmental models can fully capture the trends in Ontario's COVID-19 case counts during the investigated period. Of the five models considered, the proposed vaccination-stratified SEPAIQRD provides the closest fit to the data, both for total and stratified daily case counts (lowest RMSE in each column of Tables \ref{tab:rmse_150}, \ref{tab:rmse_64}, \ref{tab:rmse_86}). Visually, it is the only model with estimated trajectories that can partially capture the resurgence of cases in late March, and almost all actual counts lie within the 95\% credible bands in Figure \ref{fig:my_model}.

The simple SIR and SEIRD models provide similar RMSEs, performing relatively well among the models considered. However, they cannot capture the resurgence of cases that occurs in late March. The limited number of model parameters only allow them to fit the general downward trend of case counts, and they lack the flexibility to model more complex scenarios, e.g., multiple waves of the epidemic within the investigated period. The additional compartments `E' and `D' introduced in the SEIRD model do not provide the capacity to help in that regard. While these two models do not explicitly account for different vaccination statuses, simply allocating the estimated cases according to the proportion of the population in each strata provides reasonable results.

The vaccination-stratified SIR model does not perform well on this dataset, having the largest RMSEs overall of the models considered. This might be attributed to the stringent assumptions employed by its authors. First, they encoded assumptions on the efficacy of the vaccine, such that 80$\%$ of vaccinated individuals have permanent immunity, while 20$\%$ of unvaccinated individuals are assumed to be immune. 
Second, values of the model parameters were obtained by authors from existing literature, without proposing a calibration process from real data. Simply reusing their parameter values does not provide an adequate fit over our investigated period. Thus, while stratifying by vaccination status could potentially provide more granularity for predictions, the performance of the model is hindered by its fixed parameters.

The $\mathrm{SV^2(AIR)^3}$ model incorporates the impact of vaccine efficacy, policy measures, and clinical characteristics of specific COVID-19 variants. We updated the model parameters according to the characteristics of the Omicron variant, and used the actual values of the Oxford Stringency Index over the investigated period. 
However, we were unable to calibrate model trajectories that fit the data well: unvaccinated cases show an increasing rate of growth from mid-March to the end of our investigated period (Figure \ref{fig:layton_cmaj}), which is opposite to the trend in the actual data. Furthermore, while we adjusted the vaccine efficacy against the Omicron variant to be only 30\% when fully vaccinated (compared to the authors' original assumption of 75\% for the hypothetical variant), the model still vastly underestimates the number of vaccinated cases. Only towards the end of the fifth reopening phase, with the loosest restrictions, do the model trajectories start to show an uptick in vaccinated cases.
Thus, while the $\mathrm{SV^2(AIR)^3}$ model should theoretically have the flexibility to capture complex transmission dynamics, a more sophisticated method of calibrating its parameters would likely be needed to adapt it to the present setting.
 
The proposed vaccination-stratified SEPAIQRD model extended an existing SEAPIR model, by incorporating four vaccination statuses and adding other relevant compartments. We used a mix of parameters from the existing literature, together with Bayesian statistical methods to calibrate a selective set of parameters pertaining to asymptomatic infection and case ascertainment rates. For the investigated period, this approach provided a good balance between modeling flexibility and fixed parameter assumptions, with good empirical performance relative to the other models considered.

\subsection*{Examining the Calibrated Parameters}

Next, we discuss the values of the fixed and calibrated parameter values in each model. The parameters of the SIR, SEIRD and vaccination-stratified SIR model are presented in Table \ref{tab:model_para}. First, we find that calibrating parameters for the SIR and SEIRD models on case counts alone cannot accurately describe the clinical characteristics of COVID-19.
In obtaining the parameters that provide the best fit to the data for the investigated period, these two models tend to underestimate the transmission rate $\beta$ and the basic reproduction number $R_0$. Interestingly, although the SEIRD model could not calibrate a reasonable value for $\rho$ (as death counts were not used) and its estimated $R_0$ differs significantly from the SIR model, both models effectively provided the same quality of fit to the data. This suggests that the calibrated parameter values of these models should be interpreted with caution, and do not necessarily correspond to the actual clinical characteristics of the disease. The low $R_0$ values are a clear artifact of reasonably fitting the overall downward trend in case counts during the investigated period. Overall, the simplicity of these models is both a strength and a weakness. In contrast, the vaccination-stratified SIR model used entirely fixed parameters \cite{fisman2022impact}, as shown in the corresponding row of Table \ref{tab:model_para}. While its fixed $R_0$ value might more closely reflect the intrinsic spread of COVID-19, real-world factors during the investigated period violated that assumption.

\begin{table}[hbt!]
\caption{Summary of parameters in the SIR, SEIRD, and vaccination-stratified SIR models.}
\begin{center}
\begin{tabular}{@{}lcccccccc@{}}
\toprule
\multicolumn{1}{c}{\multirow{2}{*}{Model}} & \multicolumn{8}{c}{Parameter}                                                                                                                                                                                                      \\ \cmidrule(l){2-9} 
\multicolumn{1}{c}{}                       & \multicolumn{1}{l}{$\beta$} & \multicolumn{1}{l}{$\gamma$} & \multicolumn{1}{l}{$\rho$} & \multicolumn{1}{l}{$\mu$} & \multicolumn{1}{l}{$R_0^{~1}$} & \multicolumn{1}{l}{$\eta^{~2}$ } & \multicolumn{1}{l}{$PV^{~3}$ } & \multicolumn{1}{l}{$VE^{~4}$} \\ \midrule
SIR                                        & 0.0510                      & 0.0736                       & -                          & -                         & 0.6933${~}^5$                   & -                          & -                        & -                        \\
SEIRD                                      & 0.4426                      & 0.0321                       & 0.9995                     & 0.9995                    & 0.2117$^{~6}$                    & -                          & -                        & -                        \\
Vaccination-stratified SIR                 & 437                         & 73                           & -                          & -                         & 6                         & 0.5                        & 0.8                      & 0.8                      \\ \bottomrule
\end{tabular}
\end{center}
\footnotesize

Notes: \\
${~}^1 $ $R_0$ is the basic reproduction number, which governs the rate of disease spread \\
${~}^2 $ Mixing parameter between vaccinated groups and unvaccinated groups \\
${~}^3 $ Proportion of population vaccinated \\
${~}^4 $ Vaccine effectiveness, i.e., proportion of vaccinated population that is immune \\
${~}^5 $ $R_0$ in SIR model is calculated by $\frac{\beta}{\gamma}$ \\
${~}^6 $ $R_0$ in SEIRD model is calculated by $\frac{\beta}{\rho + \mu} \cdot \frac{S_0}{N}$

\label{tab:model_para}
\end{table}

The full list of model parameters in the $\mathrm{SV^2(AIR)^3}$ model is presented in Table \ref{tab:para_sav_alpha} and \ref{tab: para_sav_delta} in the Supplementary Information. The parameters we used for the emerging variant that were calibrated to values that reflect our best knowledge of the Omicron variant. These include a higher asymptomatic proportion (60\%, vs.~50\% for previous variants), a higher baseline transmission rate (4.5 times that of Delta in the unvaccinated population), a shorter recovery time (8 days), and setting the actual start date for its spread in Ontario to be November 22, 2021. We also greatly increased Omicron's transmission rates in the fully vaccinated population to reflect lower vaccine efficacy: 70\% of the baseline unvaccinated rate (compared to the authors' 12\% for Delta). Despite these calibrations, the model could not adequately describe the data during the investigated period, especially for the vaccinated population. This indicates that other assumptions used throughout the model may also require adjustment, such as the parameters related to waning immunity from vaccination.

\begin{figure}
    \centering
    \includegraphics[width = \textwidth]{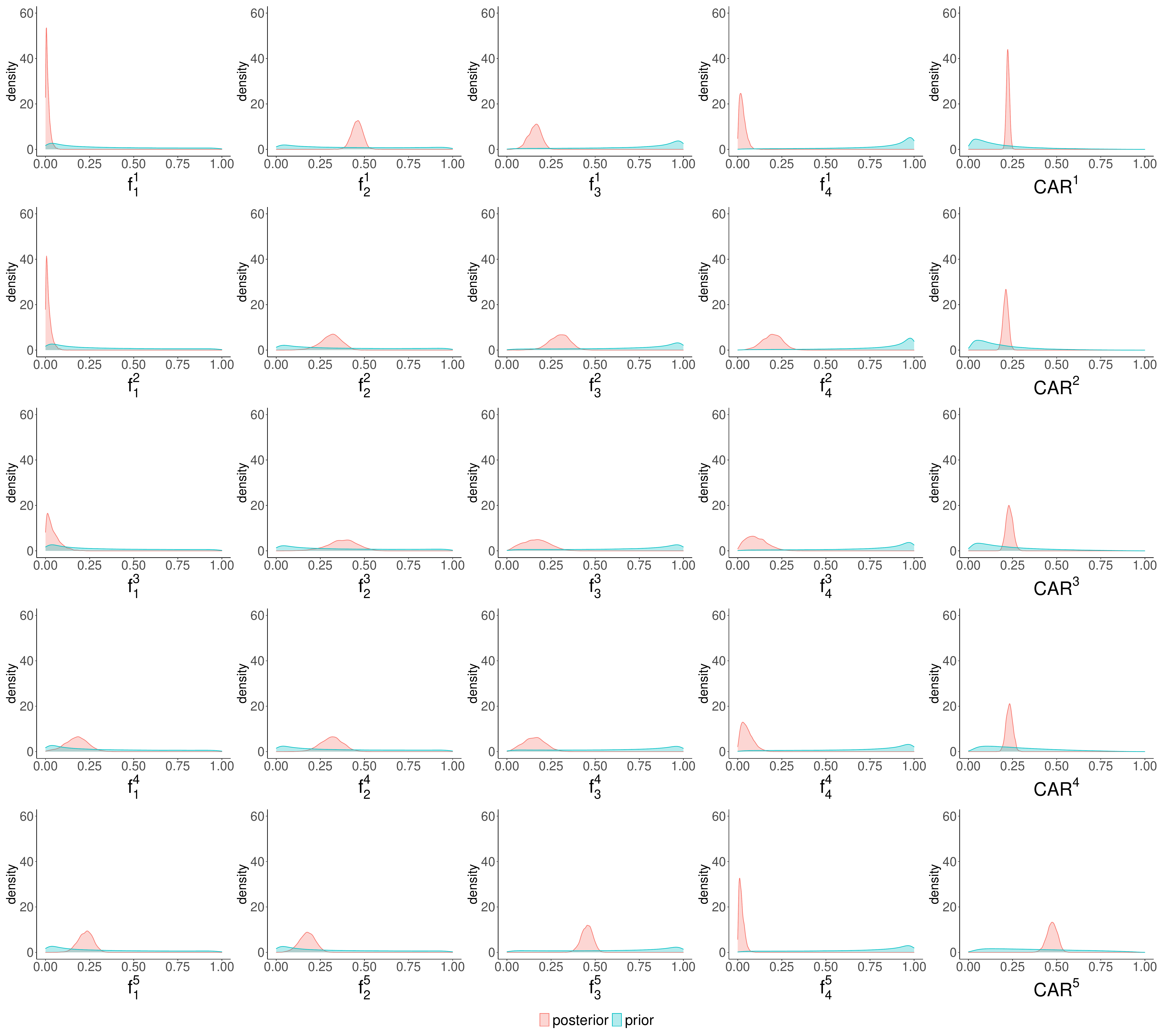}
    \caption{Comparison of prior and posterior probability densities for the fraction of asymptomatic
infection and case ascertainment rate. Rows correspond to the five reopening phases, and the first four columns correspond to the four vaccination statuses in the vaccination-stratified SEPAIQRD model: unvaccinated, partially vaccinated, completely vaccinated, and vaccinated with additional booster.}
    \label{fig:my_model_para}
\end{figure}

In the vaccination-stratified SEPAIQRD model, the calibrated parameters (i.e., $f_i^j$ and $CAR^j$) are obtained by Bayesian inference and MCMC. All four MCMC chains are observed to have converged, as shown in Figure \ref{fig:traceplot} in the Supplementary Information. A comparison between the prior and posterior probability densities of $f_i^j$ and $CAR^j$ for the five phases is plotted in Figure \ref{fig:my_model_para}. A corresponding summary table of the posterior mean, 0.025 lower quantile, and 0.975 upper quantile of the $95\%$ credible bounds for $f_i^j$ and $CAR^j$ 
is presented in Table \ref{tab:propose_para} of the Supplementary Information.
In general, the posterior means of the asymptomatic infection proportion ($f_i^j$) among all four vaccination statuses are very small. Based on the model, we might conclude it is highly likely that exposed populations become infected with at least mild symptoms, i.e., asymptomatic infection is a low-probability event according to the model. However, this cannot be fully tested against reality and could be an artifact of the model setup. The posterior means of $CAR^i$ generally increase when the Ontario government shifts from one phase to another, which implies the Ontario government is more efficient at documenting the infections, or a larger proportion of infected people get tested, when daily infections become fewer.

The credible intervals of the $f_i^j$'s broadly overlap. This implies that according to the fitted model, there is no observable difference in their posterior distributions across different vaccination statuses.
This suggests the asymptomatic infection proportion is not directly associated with the status of vaccination and the change of reopening phase, despite the prior beliefs (blue densities in Figure \ref{fig:my_model_para}) encoded in the model.
Other confounding factors might also significantly influence the asymptomatic infection proportion. For example, patients with higher-risk medical histories, such as hypertension and chronic obstructive pulmonary disease, were given priority for booster doses. Even with a booster dose, this group is increasingly likely to be infected with symptoms \cite{niu2020clinical}. Furthermore, as the Ontario government continued to relax social restrictions, the overall increase in social interactions could have a more adverse effect on highly susceptible populations. Finally, case ascertainment rates may not be uniform across different vaccination statuses.

In contrast, the calibrated values of $CAR^i$ are in good agreement with our prior beliefs. It implies that these parameters are associated with the changes in reopening phases. Since Ontario's testing policies did not change during our investigated period, a possible reason is that with the very high daily infections in phase one, it might have been difficult for the Ontario government to handle the testing volume and people were less likely to get tested.

\section*{Conclusion}
It is necessary to collect, analyze and monitor pandemic data to assess strategies of intervention, management, and control \cite{abolmaali2021comparative}. This paper aimed to provide insight into the data analysis step, by presenting a comparative study of five compartmental models and their ability to fit COVID-19 case data in Ontario, Canada from January 2022 to June 2022. In addition to four existing compartmental models, we presented an extension of the SEAPIR model to help provide a more comprehensive description of the recent COVID-19 dynamics in Ontario.
Each model was found to have its strengths and weaknesses when applied to the investigated period. The SIR and SEIRD models had relatively few compartments and simple assumptions, which allowed them to fit the overall downward trend in cases -- but not to reflect more complex situations involving multiple epidemic waves, nor necessarily have calibrated parameter values that reflect actual clinical characteristics of COVID-19. The trajectories of the vaccination-stratified SIR model and the $\mathrm{SV^2(AIR)^3}$ model appeared to be implausible compared to the actual case counts, despite them being more sophisticated models. Their implausibility and underperformance might be due to having some fixed parameters borrowed from existing literature that were no longer appropriate. Due to the real-world complexities underlying the current Ontario data, more data-driven parameters would be needed to account for situations such as time-varying case ascertainment rates and vaccine efficacy. These results practically illustrate the potential tradeoffs between applying simple models versus more complex ones.
Relative to the other models, the proposed vaccination-stratified SEPAIQRD model calibrated by Bayesian statistical methods provided the most reasonable results on this dataset, for both the estimated daily confirmed case counts and the interpretations of the calibrated parameter values.

Several limitations also exist in our work. On one hand, all model estimates are symptomatic infections. Although the assumption that Public Health Ontario only documents the number of symptomatic infections might be reasonable, asymptomatic infection is still worth consideration. At worst, the ``infected'' here is some combination of both symptomatic and asymptomatic infections, with the symptomatic very likely being the larger component.  Had the infected been separated out in the data into symptomatic and asymptomatic components, this could have been incorporated into the model (though likely the asymptomatic would be under-represented in the data).
Second, as with any statistical model, the predictive capacity has not (as yet) been tested on future case counts.  It could very well perform poorly on future counts, especially should the dynamics of disease transmission and health policy change. What is clear from this study, is that the demonstrable failures and inherent limitations of compartmental models suggest that they should not be relied on too heavily by decision-makers in forming public health policy on COVID-19.

There are several extensions of our current work that can be considered for further studies. The literature on compartmental modeling for COVID-19 transmission dynamics is vast.  Additional models, including time series models (e.g., ARIMA and SARIMA) might be considered and compared with those considered in this study. Data from other time periods or jurisdictions could also be investigated.
Finally, while Bayesian parameter calibration via MCMC methods is effective for obtaining credible bounds for parameters and estimated case counts, it comes with a relatively large computational cost. Faster computational methods for Bayesian inference would be useful for larger studies involving compartmental models.

\section*{Data Availability}
The computer code produced in this study for the proposed vaccination-stratified SEPAIQRD model is available in \url{https://github.com/YuxuanZhao1/Code-for-Vaccination-stratified-SEPAIQRD-model}. The datasets analysed during the current study are available in the Public Health Ontario repository,  \url{https://data.ontario.ca/en/dataset/covid-19-vaccine-data-in-ontario/}.

\bibliography{bibfile}

\section*{Acknowledgements}

We thank Wayne Oldford for constructive comments on the manuscript. This work was partially supported by Discovery Grant RGPIN-2019-04771 from the Natural Sciences and Engineering Research Council of Canada.




\newpage

\appendix

\renewcommand\thefigure{S\arabic{figure}}    
\setcounter{figure}{0}  

\renewcommand\thetable{S\arabic{table}}    
\setcounter{table}{0}  

\renewcommand\theequation{S\arabic{equation}}    
\setcounter{equation}{0}  

\section*{Supplementary Information}

\subsection*{A. $\mathrm{SV^2(AIR)^3}$ model}
Following Layton and Sadria \cite{layton2022understanding}, the model is initialized on January 1, 2020, and simulates the emergence of wild-type, Alpha-type, Delta-type until Fall 2021. From November 22, 2021 onwards, a new variant (which we have updated to mimic Omicron-type characteristics) replaces wild-type. We extracted the model output for our investigated period of January 6, 2022 to June 4, 2022.

\begin{table}[hbt!]
\caption{Parameters in the $\mathrm{SV^2(AIR)^3}$ model for the wild-type, alpha-type and Delta-type variants as provided in Layton and Sadria \cite{layton2022understanding}. Definitions of each model parameter are provided below, and a full description of the model can be found in Layton and Sadria \cite{layton2022understanding}.}

\begin{center}

\begin{tabular}{llll}

\toprule
\multicolumn{1}{c}{\multirow{2}{*}{Parameters Included}} & \multicolumn{2}{c}{Variant of Concern($X$)}                               \\ \cline{2-4} 

\multicolumn{1}{c}{}                                     & X=Wild               & X=Alpha              & X=Delta              \\ \hline
$\beta^X$ ${~}^1$                                          & 0.0481            & 0.0801          & 0.1107            \\
$\beta_{V1}^X$  ${~}^2$                                          & 0.0096             & 0.0401             & 0.0742             \\
$\beta_{V2=PZ}^X$   ${~}^3$                                      & 0.0024             & 0.0056             & 0.0133             \\
$\beta_{V2=AZ}^X$     ${~}^4$                                    & 0.0024             & 0.0272            & 0.0433            \\
$\beta_{R}^X$       ${~}^5$                                      & 0.0024             & 0.0040             & 0.0055             \\
$\alpha^X$        ${~}^6$                                        & 3                  & 3                  & 3                  \\
$\eta_{V1}^X$         ${~}^7$                                       & $2\cdot 0.25/182$  & $2\cdot 0.25/182$  & $2\cdot 0.25/182$  \\
$\eta_{V2= PZ,AZ}^X$      ${~}^8$                                          & $2\cdot 0.125/365$ & $2\cdot 0.125/365$ & $2\cdot 0.125/365$ \\
$\eta_R$ ${~}^9$ &     $2\cdot 0.125/365$ & $2\cdot 0.125/365$ & $2\cdot 0.125/365$ \\
$\mu$       ${~}^{10}$                                              & 0.00002            & 0.00002            & 0.00002            \\
$\mu^X$       ${~}^{11}$                                          & 0.0010             & 0.0017             & 0.0019             \\
$\mu^X_V$     ${~}^{12}$                                                & 0.00015            & 0.000255          & 0.000285           \\
$\gamma^X$     ${~}^{13}$                                             & 1/28               & 1/28               & 1/28               \\
$\sigma^X$     ${~}^{14}$                                             & 0.5                & 0.5                & 0.5                \\
$\sigma_V^X$    ${~}^{15}$                                            & 0.85               & 0.85               & 0.85               \\
$\sigma_R^X$       ${~}^{16}$                                         & 0.85               & 0.85               & 0.85               \\
 \bottomrule
\end{tabular}
\end{center}

\footnotesize
Notes: \\
${~}^1 $ Disease transmission rate without vaccination \\
${~}^2 $ Disease transmission rate after partial vaccination\\
${~}^3 $ Disease transmission rate after taking full dose of Pfizer-BioNTech or Moderna \\
${~}^4 $ Disease transmission rate after taking full dose of Astra-Zeneca\\
${~}^5 $ Disease re-infection rate\\
${~}^6 $ Ratio between asymptomatic and symptomatic infectivity\\
${~}^7 $ Loss of immunity rate after partial vaccination\\
${~}^8 $ Loss of immunity rate after full vaccination\\
${~}^9 $ Loss of immunity rate after recovering from previous infection\\
${~}^{10} $ Natural death rate\\
${~}^{11} $ Disease mortality rate without vaccination \\
${~}^{12} $ Disease mortality rate after vaccination\\
${~}^{13} $ Disease recovery rate\\
${~}^{14} $ Fraction of asymptomatic infections without vaccination\\
${~}^{15} $ Fraction of asymptomatic infections after vaccination \\
${~}^{16} $ Fraction of asymptomatic infections after recovering from previous infection \\

\label{tab:para_sav_alpha}
\end{table}

\begin{table}[hbt!]

\caption{Parameters in the $\mathrm{SV^2(AIR)^3}$ model for the newly-emerging variant \cite{layton2022understanding} and our updated values for the Omicron variant. Parameter definitions are the same as in Table S1. \strut}

\centering

\label{tab: para_sav_delta} 
\begin{tabular}{lll}
\toprule
\multicolumn{1}{c}{\multirow{2}{*}{Parameters Included}} & \multicolumn{2}{c}{Variant of Concern($X$)} \\ \cline{2-3} 
\multicolumn{1}{c}{}      &                                           X= Omicron & X=   Hypothetical Variant                \\ \hline
$\beta^X$                       &     0.5000                        & 0.1262 \\ 
$\beta_{V1}^X$                   &    0.4250                         & 0.0883 \\ 
$\beta_{V2 = PZ}^X$            &      0.3500                            & 0.0315 \\
$\beta_{V2 = AZ}^X$            &      0.3500                             & 0.0631\\

$\beta_{R}^X$                     &    0.0250                   & 0.0063 \\
$\alpha^X$                       &       3                          & 3                  \\
$\eta_{V1}^X$  & $2\cdot 0.25/182$ &$2\cdot 0.25/182$ \\
$\eta_{V2 = PZ,AZ}^X$  & $2\cdot 0.125/365$&$2\cdot 0.125/365$ \\
$\eta_{R}$  &$2\cdot 0.125/365$ &$2\cdot 0.125/365$ \\
$\mu$ &0.00002 & 0.00002 \\
$\mu^X$ & 0.0019& 0.0010 \\
$\mu_{V}^X$                       &          0.000284                       & 0.00015 \\
$\gamma^X$ & 1/8 & 1/28\\
$\sigma^X$ & 0.6 & 0.55\\
$\sigma^X_V$ & 0.85 & 0.85\\
$\sigma^X_R$ &0.85 & 0.85 \\

\bottomrule
\end{tabular}
\label{tab: para_sav_delta} 
\end{table}
\clearpage
\newpage
\subsection*{B. Vaccination-stratified SEPAIQRD model }
\subsubsection*{B.1 Plot of daily confirmed cases after data processing step}

\begin{figure}[hbt!]
    \centering
    \includegraphics[width = 0.75 \textwidth]{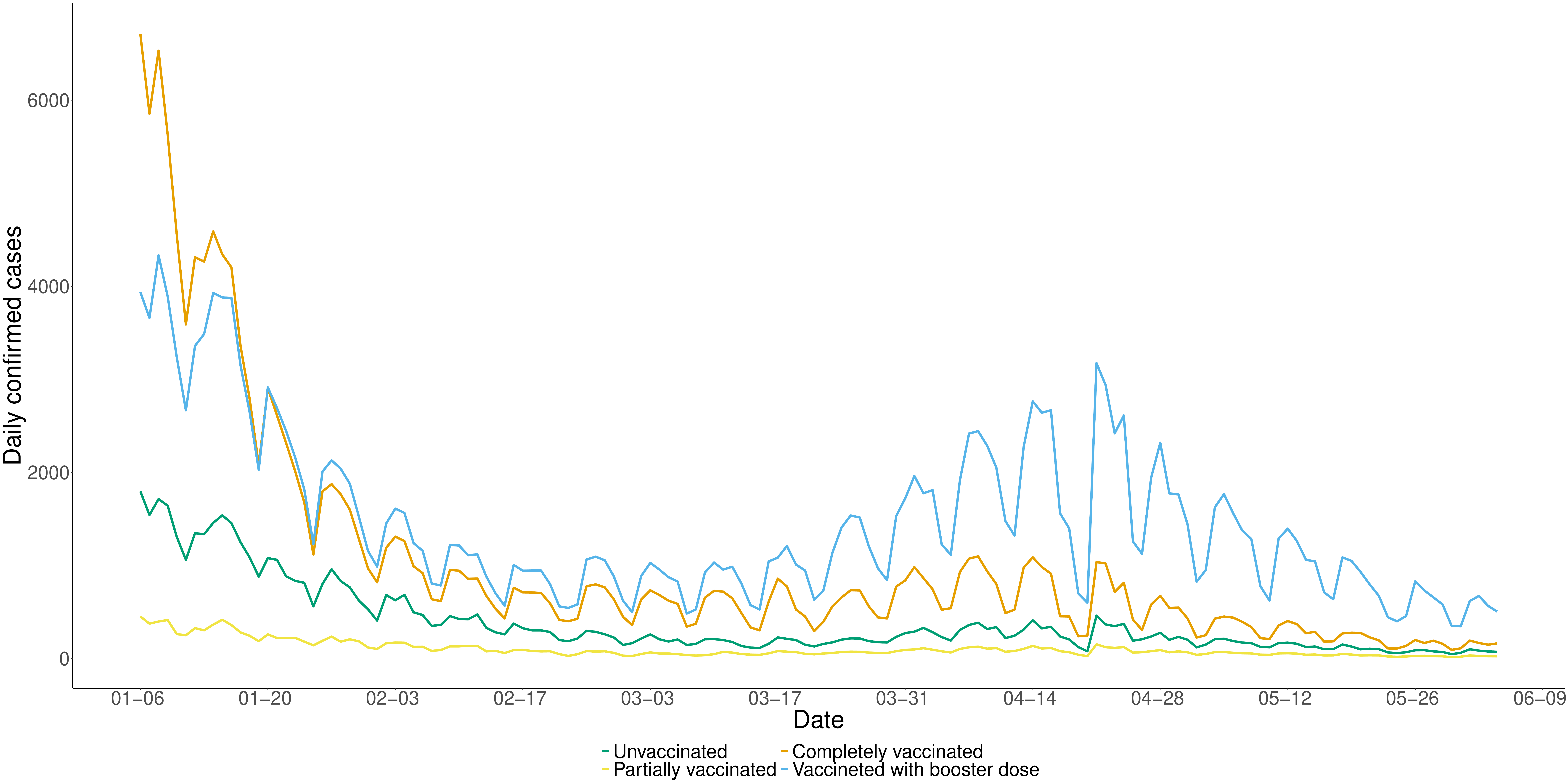}
    \caption{Daily confirmed COVID-19 cases from January 6, 2022 to June 4, 2022 in Ontario stratified by four vaccination statuses: unvaccinated, partially vaccinated, completely vaccinated, and vaccinated with booster dose. From January 6 to March 10, we split Ontario's `fully vaccinated' infections according to the proportion of completely vaccinated and vaccinated with booster dose populations in Ontario. After March 10, we split the `not fully vaccinated' infections according to the daily-updated proportion of unvaccinated population and partially vaccinated population in Ontario.}
    \label{fig:str}
\end{figure}
\subsubsection*{B.2 Detailed description of the disease transmission stage}

In the disease transmission stage, the interactions within the same group and between different groups are considered. We follow the idea from existing literature \cite{FIELDS2021e07905} to quantify the transmission rate between stratified groups using 
$$
    \beta_{ij} = c_{ij} \cdot p,
$$
 where $p$ is defined as the transmission probability per contagious contact and $c_{ij}$ is defined as the daily contact rate between the i$^\mathrm{th}$ and j$^\mathrm{th}$ groups.
To the best of our knowledge, there are not any available data evaluating the contact rate between groups with different vaccination statuses. We take the idea of the contact matrix given age \cite{FIELDS2021e07905}, and apply it to construct a contact matrix given vaccination status assuming independence. The transformed contact matrix is shown in Table \ref{tab:contact}. Then, we set $p = 0.02$ \cite{abdollahi2020simulating}.

 \begin{table}[hbt!]
 \caption{Contact matrix given groups with different vaccination statuses. Note that $c_{ij}\ne c_{ji}$ since the case where people in the i$^\mathrm{th}$ group actively contact people in the j$^\mathrm{th}$ group is distinct from the case where people in the j$^\mathrm{th}$ group actively contact people in the i$^\mathrm{th}$ group.}
\centering
\resizebox{\textwidth}{!}{
\begin{tabular}{@{}llllll@{}}
\toprule
                                      &                         &              & \multicolumn{2}{c}{Passive Contact Group(j)}    &                         \\ \midrule
                                      &                         & Unvaccinated & Partially vaccinated & Completely vaccinated & Vaccinated with booster \\
\multirow{4}{*}{Active Contact Group(i)} & Unvaccinated            & 2.4844       & 1.1046               & 6.3798                & 4.1864                  \\
                                      & Partially vaccinated    & 2.7819       & 1.3353               & 6.3711                & 3.1741                  \\
                                      & Completely vaccinated   & 1.8744       & 0.7432               & 5.7009                & 5.3397                  \\
                                      & Vaccinated with booster & 0.9392       & 0.2827               & 4.0765                & 6.3317                  \\ \bottomrule
\end{tabular}}
\label{tab:contact}
\end{table}

The individuals who are considered contagious include those in the documented symptomatic infectious compartment ($I^i$), symptomatic infectious compartment ($P^i$), asymptomatic infectious compartment ($A^i$), and two transition compartments ($T^i$ and $T'^i$). Different compartments may transmit the disease at different rates: we let $\beta_{ij}^I, \beta_{ij}^P, \beta_{ij}^A, \beta_{ij}^T, \beta_{ij}^{T'}$, respectively denote the transmission rates of these contagious compartments. We assume people in $P^i$, $T^i$, and $T'^i$ are as contagious as those in $I^i$, such that \begin{equation}
    \beta_{ij}^P =\beta_{ij}^T =\beta_{ij}^{T'} = \beta_{ij}^I = \beta_{ij} = c_{ij}\cdot p.\label{eqn:b_ijp}\end{equation} 
We also assume those in $A^i$ are not as contagious as those in $I^i$, such that
\begin{equation}
\beta_{ij}^A = 0.2\cdot \beta_{ij} =0.2\cdot  c_{ij}\cdot p.
\label{eqn:b_ija}
\end{equation}
With the setup  above, we can compute the disease transmission rates of $I^i$,$P^i$ and $A^i$  using Equation \ref{eqn:b_ijp} and \ref{eqn:b_ija}. Table \ref{tab:contact_IP} shows the disease transmission rates of the $I^i$ and $P^i$ compartments, while Table \ref{tab:contact_IA} shows the disease transmission rate of compartment  $A^i$.

  \begin{table}[hbt!]
 \caption{Disease transmission rates of the $I^i$, $P^i$, $T^i$, and $T'^i$ compartments, which are respectively denoted as $\beta_{ij}^I$, $\beta_{ij}^P$, $\beta_{ij}^T$, and $\beta_{ij}^{T'}$}
\centering
\resizebox{\textwidth}{!}{
\begin{tabular}{@{}llllll@{}}
\toprule
                                      &                         &              & \multicolumn{2}{c}{Passive Contact Group(j)}    &                         \\ \midrule
                                      &                         & Unvaccinated & Partially vaccinated & Completely vaccinated & Vaccinated with booster \\
\multirow{4}{*}{Active Contact Group(i)} & Unvaccinated            & 0.0497       & 0.0221            & 0.1276                & 0.0837               \\
                                      & Partially vaccinated    & 0.0556      & 0.0267              & 0.1267                & 0.0635                  \\
                                      & Completely vaccinated   & 0.0375    & 0.0149               & 0.1140             & 0.1068                  \\
                                      & Vaccinated with booster & 0.0188       & 0.0057              & 0.0815                & 0.1266                 \\ \bottomrule
\end{tabular}}
\label{tab:contact_IP}
\end{table}

   \begin{table}[hbt!]
 \caption{Disease transmission rate of the $A^i$ compartment, which is denoted as $\beta_{ij}^A$}
\centering
\resizebox{\textwidth}{!}{
\begin{tabular}{@{}llllll@{}}
\toprule
                                      &                         &              & \multicolumn{2}{c}{Passive Contact Group(j)}    &                         \\ \midrule
                                      &                         & Unvaccinated & Partially vaccinated & Completely vaccinated & Vaccinated with booster \\
\multirow{4}{*}{Active Contact Group(i)} & Unvaccinated            & 0.0099      & 0.0044            & 0.0255             & 0.0167              \\
                                      & Partially vaccinated    & 0.0111    & 0.0053             & 0.0253                & 0.0127                  \\
                                      & Completely vaccinated   & 0.0750    & 0.0030              & 0.0228             & 0.0214                \\
                                      & Vaccinated with booster & 0.0038       & 0.0011              & 0.0163                & 0.0253                \\ \bottomrule
\end{tabular}}
\label{tab:contact_IA}
\end{table}
The impact of policy changes on COVID-19 transmission is also considered. We use the time-varying Oxford Stringency Index to quantify this impact. We scale our $\beta_{ij}$ by the time-dependent $1-\lambda(t)$. This is a piece-wise constant function, which means that it will remain constant within the same reopening phase and changes as the Ontario government moves from one phase to another. The changepoints used in the definition of $ 1-\lambda(t)$ align with the changes in reopening states described in the Data Description section of the main paper.
\clearpage
\newpage

\subsubsection*{B.3 Fixed model parameters}
Existing literature provides estimates of the case fatality proportion by vaccination status, which need to be converted to a death rate with the unit of days$^{-1}$. Let $\zeta_i$ denote the case fatality proportion for vaccination status $i$. The corresponding death rate $\alpha_i$ can be converted by $$\frac{\zeta_i}{\zeta_i + (1-\zeta_i)\cdot\kappa_{I\to Q}}.$$ Since the self-isolation delay $\kappa_{I\to Q}$ is set to 1 day, the death proportion $\zeta_i$ equals the death rate $\alpha_i$.  The self-isolation compliance proportion $\epsilon$ is assumed to be 0.96. Sources for the other parameters are indicated in Table 2 of the main text.

\begin{table}[hbt!]
 \centering
 \caption{Values of fixed parameters in the vaccination-stratified SEPAIQRD model. \strut}
\begin{tabular}{@{}ll@{}}
\toprule
Model Parameter   & Value             \\ \midrule
$\kappa_E$        & 1/3.1 days$^{-1}$ \\
$\kappa_A$        & 1/7.6 days$^{-1}$ \\
$\kappa_{I\to Q}$ & 1 days$^{-1}$   \\
$\kappa_{Q\to R}$ & 1/4.7 days$^{-1}$ \\
$\kappa_{I\to R}$ & 1/5.7 days$^{-1}$ \\
$\kappa_P$        & 1/2.1 days$^{-1}$     \\
$\kappa_{P\to R'}$       & 1/7.7 days$^{-1}$ \\
$\epsilon$       & 0.96 \\
$\alpha_1$        & 0.79/100    days$^{-1}$      \\
$\alpha_2$        & 0.7/100 days$^{-1}$      \\
$\alpha_3$        & 0.14/100 days$^{-1}$         \\
$\alpha_4$        & 0.12/100 days$^{-1}$     \\ \bottomrule
\end{tabular}
\label{tab:parameters_proposed}
\end{table}

\clearpage
\newpage
\subsubsection*{B.4 System of differential equations }

This section provides the system of differential equations that governs the dynamic mechanisms of our vaccination-stratified SEPAIQRD model, corresponding to the schematic in Figure 2 of the main text.

We define $N^1,N^2,N^3,N^4$ to be piecewise-constant functions varying by day, with values corresponding to the sub-population sizes of the four vaccination statuses. Likewise, we define $V^1,V^2,V^3$ to be piecewise-constant functions varying by day, which are given by the number of first, second, and third doses given to individuals per day. Both $N^i$ and $V^j$ are directly calculated from the COVID-19 vaccination data from Public Health Ontario. Then, the differential equations at the disease transmission stage for the four susceptible subpopulations can be written as follows:
$$
     \frac{d S^{1 }}{d t}=-\sum_{j=1}^4\frac{S^{1 }} {N^{j }}\left(\beta_{P}^{1j}P^{j }+\beta_{I}^{1j} I^{j }+\beta_{A}^{1j}A^{j }+\beta_{T}^{1j}T^{j }+\beta_{T'}^{1j}T'^{j }\right)\cdot (1-\lambda(t))- V^1
$$
$$\frac{d S^{2 }}{d t}=-\sum_{j=1}^4\frac{S^{2 }} {N^{j }}\left(\beta_{P}^{2j}P^{j }+\beta_{I}^{2j} I^{j }+\beta_{A}^{2j}A^{j }+\beta_{T}^{2j}T^{j }+\beta_{T'}^{2j}T'^{j }\right)\cdot (1-\lambda(t)) +V^1 - V^2
$$
 $$\frac{d S^{3 }}{d t}=-\sum_{j=1}^4\frac{S^{3 }} {N^{j }}\left(\beta_{P}^{3j}P^{j }+\beta_{I}^{3j} I^{j }+\beta_{A}^{3j}A^{j }+\beta_{T}^{3j}T^{j }+\beta_{T'}^{3j}T'^{j }\right)\cdot (1-\lambda(t))+V^2 - V^3
$$
$$\frac{d S^{4 }}{d t}=-\sum_{j=1}^4\frac{S^{4 }} {N^{j }}\left(\beta_{P}^{4j}P^{j }+\beta_{I}^{4j} I^{j }+\beta_{A}^{j}A^{j }+\beta_{T}^{4j}T^{j }+\beta_{T'}^{4j}T'^{j }\right)\cdot (1-\lambda(t)) +V^3.
$$
The remaining compartments are governed by the following  differential equations. 
$$
\frac{d E^{i }}{d t}=\sum_{j=1}^4\frac{S^{i }} {N^{j }}\left(\beta_{P}^{ij}P^{j }+\beta_{I}^{ij} I^{j }+\beta_{A}^{ij}A^{j }+\beta_{T}^{ij}T^{j }+\beta_{T'}^{ij}T'^{j }\right)\cdot (1-\lambda(t))-(1-f_i(t))\cdot\kappa_{E}\cdot E^{i }-f_i(t)\cdot \kappa_E \cdot E^i 
$$$$
\frac{d A^{i }}{d t}=f_i(t)\cdot\kappa_E\cdot E^i-\kappa_{A} A^{i } 
$$$$
\frac{d RA^{i }}{d t} = \kappa_A A^{i }
$$$$
\frac{d P^{i }}{d t}=(1-f_i(t))\cdot\kappa_{E}\cdot E^i-\kappa_P\cdot CAR(t)\cdot P^i-\kappa_{P}\cdot(1-CAR(t))\cdot P^i
$$$$
\frac{d T'^{i }}{d t}=(1-CAR(t))\cdot \kappa_{P}\cdot P^i -\frac{1}{\frac{1}{\kappa_{P\to R'}}-\frac{1}{\kappa_{P}}}\cdot T'^i
$$$$
\frac{d R'^{i }}{d t}= \frac{1}{\frac{1}{\kappa_{P\to R'}}-\frac{1}{\kappa_{P}}}\cdot T'^i
$$$$
\frac{d I^{i }}{d t}=\kappa_{P}\cdot CAR(t)\cdot P^{i }-\left(\alpha_i+(1-\epsilon)\cdot(1-\alpha_i)\cdot\kappa_{I\to R}+(1-\alpha_i)\cdot\epsilon\cdot\kappa_{I\to Q}\right) \cdot I^{i } 
$$$$
\frac{d T^{i }}{d t} = (1-\alpha_i)\cdot (1-\epsilon) \cdot \kappa_{I\to Q}\cdot I^i -\frac{1}{\frac{1}{\kappa_{I\to R}}-\frac{1}{\kappa_{I\to Q}}}\cdot T^i
$$$$
\frac{d R^{i }}{d t}=\frac{1}{\frac{1}{\kappa_{I\to R}}-\frac{1}{\kappa_{I\to Q}}}\cdot T^i + \frac{1}{\frac{1}{\kappa_{I\to R}}-\frac{1}{\kappa_{I\to Q}}}\cdot Q^i
$$$$
\frac{d Q^{i }}{d t} = (\kappa_{I\to Q}\cdot \epsilon\cdot(1-\alpha_i))I^{i }-\frac{1}{\frac{1}{\kappa_{I\to R}}-\frac{1}{\kappa_{I\to Q}}}\cdot Q^i
$$$$
\frac{d D^{i }}{d t} = \alpha_i\cdot I^{i }
$$
\clearpage
\newpage
\subsubsection*{B.5 Prior distributions of unknown parameters}

 \begin{table}[hbt!]
\centering
\caption{The prior distributions of the unknown parameters $\mathcal{L}(f_i^j)$,  $\mathcal{L}(CAR^j)$, and $(\phi_i^j)^{-1}$. The superscript $i$ indicates the vaccination status, and the subscript $j$ indicates the reopening phase. As described in the main text, the prior means for $\mathcal{L}(CAR^j)$ are set to increase as Ontario moves through reopening phases. The prior means for the logit-transformed asymptomatic infection proportion are set by vaccination status and reopening phase, such that vaccination reduces symptoms (more likely to be asymptomatic with more doses) and efficacy of vaccination decays over the investigated period. Large prior standard deviations are set so that the posterior distributions will be primarily informed by the data. \strut}
\resizebox{4.5cm}{!}{
\begin{tabular}{@{}ll@{}}
\toprule
Parameter            & Prior distribution \\ \midrule
$\mathcal{L}(f_1^1)$ & Normal(-1,2)       \\
$\mathcal{L}(f_2^1)$ & Normal(-0.5,2)     \\
$\mathcal{L}(f_3^1)$ & Normal(1.5,2)      \\
$\mathcal{L}(f_4^1)$ & Normal(2,2)        \\
\addlinespace
$\mathcal{L}(f_1^2)$ & Normal(-1,2)       \\
$\mathcal{L}(f_2^2)$ & Normal(-0.65,2)    \\
$\mathcal{L}(f_3^2)$ & Normal(1.25,2)     \\
$\mathcal{L}(f_4^2)$ & Normal(1.75,2)     \\
\addlinespace
$\mathcal{L}(f_1^3)$ & Normal(-1,2)       \\
$\mathcal{L}(f_2^3)$ & Normal(-0.75,2)    \\
$\mathcal{L}(f_3^3)$ & Normal(1,2)        \\
$\mathcal{L}(f_4^3)$ & Normal(1.5,2)      \\
\addlinespace
$\mathcal{L}(f_1^4)$ & Normal(-1,2)       \\
$\mathcal{L}(f_2^4)$ & Normal(-0.85,2)    \\
$\mathcal{L}(f_3^4)$ & Normal(0.8,2)      \\
$\mathcal{L}(f_4^4)$ & Normal(1.25,2)     \\
\addlinespace
$\mathcal{L}(f_1^5)$ & Normal(-1,2)       \\
$\mathcal{L}(f_2^5)$ & Normal(-0.95,2)    \\
$\mathcal{L}(f_3^5)$ & Normal(0.75,2)     \\
$\mathcal{L}(f_4^5)$ & Normal(1.15,2)     \\
\addlinespace
$\mathcal{L}(CAR^1)$ & Normal(-1.84,1.31) \\
$\mathcal{L}(CAR^2)$ & Normal(-1.8, 1.15) \\
$\mathcal{L}(CAR^3)$ & Normal(-1.5,1.31)  \\
$\mathcal{L}(CAR^4)$ & Normal(-1.09,1.15) \\
$\mathcal{L}(CAR^5)$ & Normal(-0.59,1.31) \\
\addlinespace
$(\phi_i^j)^{-1}$           & Exponential(5)     \\ \bottomrule
\end{tabular}}
\label{tab:prior_parameters}
\end{table}

\newpage

\subsubsection*{B.6 Posterior distributions of unknown parameters}


\begin{table}[hbt!]
\centering
\caption{Summary of posterior distributions of $CAR^j$ and $f_i^j$ on the original [0,1] scale. For each parameter, the posterior mean, lower quantile ($0.025$) of the $95\%$ credible interval, and upper quantile ($0.975$) of the $95\%$ credible interval are shown, based on the MCMC samples.}
\resizebox{!}{0.40\textwidth}
{
\begin{tabular}{lrrr}
\toprule
  & Mean & Lower Quantile & Upper Quantile\\
\midrule
$f_1^1$ & 0.0089 & 0.0009 & 0.0457\\
$f_2^1$ & 0.4578 & 0.3950 & 0.5153\\
$f_3^1$  & 0.1528 & 0.0805 & 0.2231\\
$f_4^1$ & 0.0218 & 0.0035 & 0.0710\\
\addlinespace
$f_1^2$ & 0.0117 & 0.0011 & 0.0618\\
$f_2^2$ & 0.3050 & 0.1777 & 0.4155\\
$f_3^2$ & 0.2923 & 0.1695 & 0.4022\\
$f_4^2$ & 0.1876 & 0.0785 & 0.3027\\
\addlinespace
$f_1^3$& 0.0282 & 0.0023 & 0.1430\\
$f_2^3$ & 0.3688 & 0.1916 & 0.5287\\
$f_3^3$ & 0.1489 & 0.0353 & 0.3164\\
$f_4^3$&0.0893 & 0.0148 & 0.2444\\
\addlinespace
$f_1^4$ & 0.1627 & 0.0430 & 0.2929\\
$f_2^4$ &0.3099 & 0.1879 & 0.4338\\
$f_3^4$  & 0.1425 & 0.0406 & 0.2768\\
$f_4^4$ & 0.0429 & 0.0072 & 0.1350\\
\addlinespace
$f_1^5$ & 0.2234 & 0.1388 & 0.3051\\
$f_2^5$ & 0.1697 & 0.0817 & 0.2598\\
$f_3^5$  & 0.4528 & 0.3854 & 0.5134\\
$f_4^5$ & 0.0171 & 0.0030 & 0.0584\\
\addlinespace
$CAR^1$ & 0.2238 & 0.2067 & 0.2419\\
$CAR^2$ & 0.2115 & 0.1818 & 0.2409\\
$CAR^3$ & 0.2322 & 0.1955 & 0.2731\\
$CAR^4$ & 0.2318 & 0.1964 & 0.2688\\
$CAR^5$ & 0.4736 & 0.4117 & 0.5284\\
\bottomrule
\end{tabular}
}
\label{tab:propose_para}
\end{table}


\begin{figure}[hbt!]
    \centering
    \includegraphics[width = 1.1\textwidth]{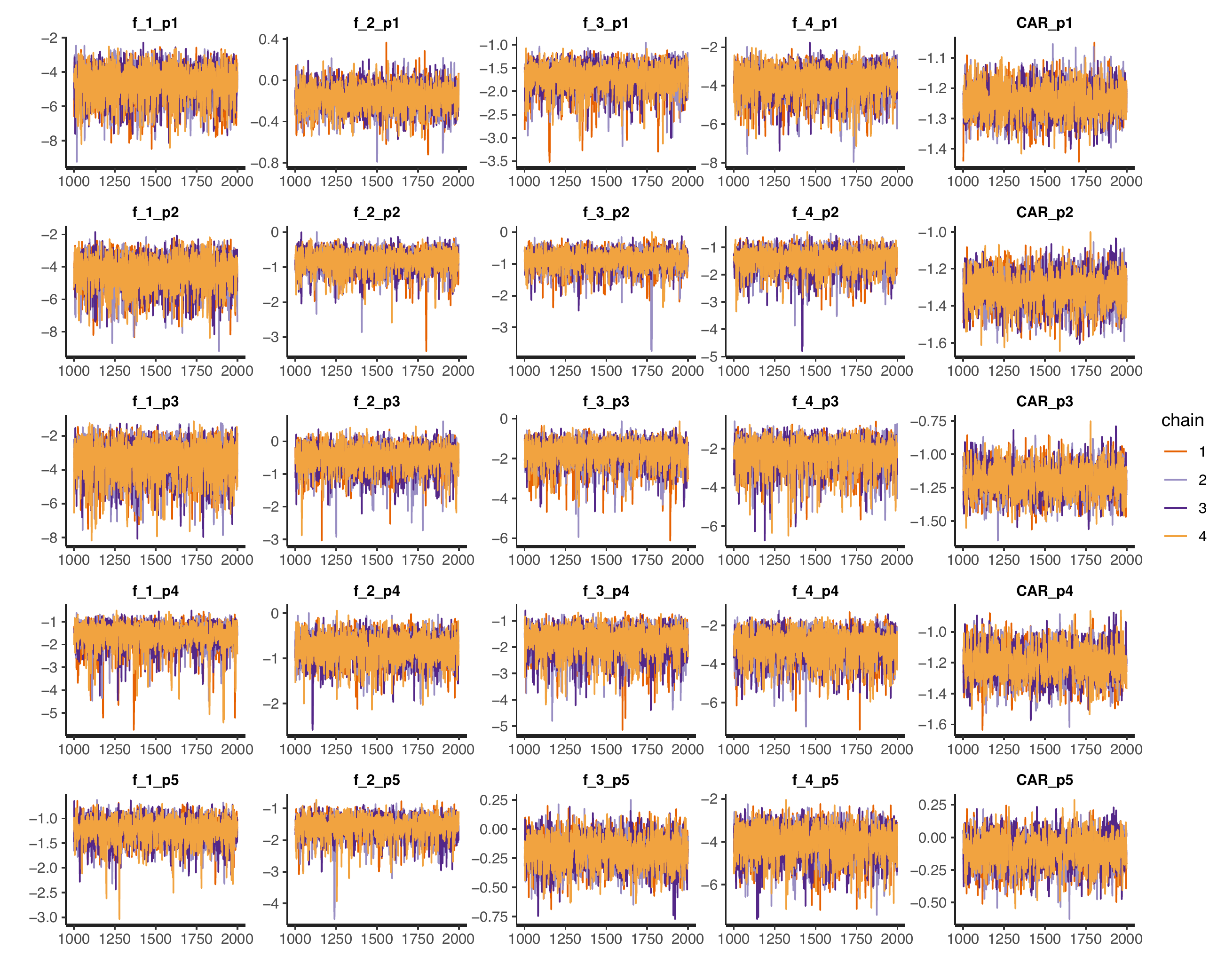}
    \caption{Traceplots of the MCMC samples for the model parameters $\mathcal{L}(f_i^j)$,  $\mathcal{L}(CAR^j)$, and $(\phi_i^j)^{-1}$.}
    \label{fig:traceplot}
\end{figure}

\end{document}